\documentclass[twocolumn]{aastex63}
\usepackage{xcolor}
\usepackage{lineno}
\usepackage[utf8]{inputenc}

\shorttitle{X-ray Phase Space of Transients}
\shortauthors{Polzin et al.}

\DeclareUnicodeCharacter{2212}{-}
\begin{document}

\title{The Luminosity Phase Space of Galactic and Extragalactic X-ray Transients Out to Intermediate Redshifts}

\correspondingauthor{Ava Polzin}
\email{apolzin@uchicago.edu}

\author[0000-0002-5283-933X]{Ava Polzin}
\affiliation{Department of Astronomy and Astrophysics, The University of Chicago, Chicago, IL 60637, USA}

\author[0000-0003-4768-7586]{Raffaella Margutti}
\affiliation{Department of Astronomy, University of California, Berkeley, CA, 94720, USA}
\affiliation{Department of Physics, University of California, Berkeley, CA, 94720, USA}

\author[0000-0001-5126-6237]{Deanne L. Coppejans}
\affiliation{Department of Physics, University of Warwick, Gibbet Hill Road, Coventry CV4 7AL, UK}
\affiliation{Department of Physics \& Astronomy and Center for Interdisciplinary Exploration and Research in Astrophysics, Northwestern University, Evanston, IL, 60208, USA}

\author[0000-0002-4449-9152]{Katie Auchettl}
\affiliation{School of Physics, The University of Melbourne, Parkville, VIC 3010, Australia}
\affiliation{ARC Centre of Excellence for All Sky Astrophysics in 3 Dimensions (ASTRO 3D)}
\affiliation{Department of Astronomy and Astrophysics, University of California, Santa Cruz, CA 95064, USA}

\author[0000-0001-5624-2613]{Kim L. Page}
\affiliation{School of Physics \& Astronomy, University of Leicester, Leicester LE1 7RH, UK}

\author[0000-0003-3902-3915]{Georgios Vasilopoulos}
\affiliation{Université de Strasbourg, CNRS, Observatoire astronomique de Strasbourg, UMR 7550, F-67000 Strasbourg, France}
\affiliation{Department of Physics, National and Kapodistrian University of Athens, University Campus Zografos, GR 15784, Athens, Greece}

\author[0000-0002-7735-5796]{Joe S. Bright}
\affiliation{Astrophysics, Department of Physics, University of Oxford, Keble Road, Oxford OX1 3RH, UK}

\author[0000-0003-4849-5092]{Paolo Esposito}
\affiliation{Scuola Universitaria Superiore IUSS Pavia, Palazzo del Broletto, piazza della Vittoria 15, 27100 Pavia, Italy}
\affiliation{INAF--Istituto di Astrofisica Spaziale e Fisica Cosmica di Milano, Via A. Corti 12, 20133 Milano, Italy}

\author[0000-0003-3734-3587]{Peter K. G. Williams}
\affiliation{Center for Astrophysics $\mid$ Harvard \& Smithsonian, 60 Garden Street, Cambridge, MA 02138-1516, USA}
\affiliation{American Astronomical Society, 1667 K Street NW, Suite 800, Washington, DC, 20006, USA}

\author[0000-0002-8286-8094]{Koji Mukai}
\affiliation{CRESST II and X-ray Astrophysics Laboratory, NASA/GSFC, Greenbelt, MD 20771, USA}
\affiliation{Department of Physics, University of Maryland, Baltimore County, 1000 Hilltop Circle, Baltimore, MD 21250, USA}

\author[0000-0002-9392-9681]{Edo Berger}
\affiliation{Center for Astrophysics $\mid$ Harvard \& Smithsonian, 60 Garden Street, Cambridge, MA 02138-1516, USA}

\begin{abstract}
We present a detailed compilation and analysis of the X-ray phase space of low- to intermediate-redshift ($ 0\le z \le 1$) transients that consolidates observed light curves (and theory where necessary) for a large variety of classes of transient/variable phenomena in the 0.3--10 keV energy band. We include gamma-ray burst afterglows, supernovae, supernova shock breakouts and shocks interacting with the environment, tidal disruption events and active galactic nuclei, fast blue optical transients, cataclysmic variables, magnetar flares/outbursts and fast radio bursts, cool stellar flares, X-ray binary outbursts, and ultraluminous X-ray sources. Our overarching goal is to offer a comprehensive resource for the examination of these ephemeral events, extending the X-ray duration-luminosity phase space (DLPS) to show luminosity evolution. We use existing observations (both targeted and serendipitous) to characterize the behavior of various transient/variable populations. Contextualizing transient signals in the larger DLPS serves two primary purposes: to identify areas of interest (i.e., regions in the parameter space where one would expect detections, but in which observations have historically been lacking) and to provide initial qualitative guidance in classifying newly discovered transient signals. We find that while the most luminous (largely extragalactic) and least luminous (largely Galactic) part of the phase space is well-populated at $t > 0.1$ days, intermediate luminosity phenomena (L$_x = 10^{34} - 10^{42}$ erg s$^{-1}$) represent a gap in the phase space. We thus identify L$_x = 10^{34} - 10^{42}$ erg s$^{-1}$ and $t = 10^{-4} - 0.1$ days as a key discovery phase space in transient X-ray astronomy.
\end{abstract}


\section{Introduction} \label{sec:intro}
Transient and variable electromagnetic emission is often associated with the most violent events in space, like stellar explosions, stellar disruptions by supermassive black holes, or accretion-related phenomena on compact objects to name a few. Studying the timescales and intrinsic energy released by each of these phenomena often provide guidance to understand the physics that regulates the bright displays of these transients and variables. To this end, the duration-luminosity phase space (DLPS), where duration is defined as the time between the identification of an outburst and its later non-detection, has been used as a means of placing classes of transient and variable phenomena in the context of their underlying physics and constraining their outburst mechanisms.

Previous works have focused on building an observationally motivated, light curve-populated DLPS for specific wavelength regimes (e.g., \citealt{2012arXiv1202.2381K, 2015MNRAS.446.3687P, 2017ApJ...849...70V} for optical wavelengths, \citealt{2022ApJ...935...16E} for millimeter wavelengths, or \citealt{2015ApJ...806..224M} for radio wavelengths), which is facilitated by the significant volume of available data. We build on the first attempts to produce an observation-driven DLPS in the X-rays \citep[][]{2009astro2010S.278S, 2013RSPTA.37120498O} by populating the DLPS with light curves
as a comprehensive view of the low- to intermediate-redshift ($z\le1$, in order to ensure the sample of sources in the DLPS is representative of the overall demographics presented and that the intrinsic rate of such events is well understood) phase space for (observer frame) 0.3-10 keV transient and variable X-ray phenomena. This extends the use of the DLPS by showing both luminosity and time evolution of these events. The motivation behind compiling this dataset is two-pronged: to identify pristine regions of this parameter space that can be explored by future observing facilities (i.e., identification of discovery areas) and conversely, we can use the phase space location of an unknown type of transient to constrain its intrinsic nature. This dual motivation, both for characterizing the nature of observed events and for identifying discovery frontiers for the future generation of X-ray observatories, makes examination of the phase space vital.

With a number of large scale, all-sky transient surveys that have been carried out at the time of writing (e.g., SRG/eROSITA; \citealt{2021AA...647A...1P}) or are beginning in the immediate future, the DLPS can offer an initial rapid designation for observed events in tandem with targeted follow-up or before follow-up is initiated. The DLPS is also a resource for retroactive classification of transients recovered from archival data, when follow-up is potentially no longer possible, making it a valuable tool to determine object class from existing observations.

We utilize complete X-ray light curves for a variety of Galactic and extragalactic transient and variable phenomena: gamma-ray burst afterglows, supernovae, supernova shock breakouts and shocks interacting with the environment, tidal disruption events and active galactic nuclei, fast blue optical transients, cataclysmic variables, magnetar flares/outbursts and fast radio bursts, cool stellar flares, X-ray binary outbursts, and ultraluminous X-ray sources. For some classes the data are sparse, and we will instead plot peak X-ray luminosity (L$_x$) vs. duration.
In the one case where there were insufficient (\textit{confirmed}) observations, we used theory as a supplement.

We present in Section \ref{sec:data} the datasets for each of the different classes of transient and variable events, and we discuss their location within the DLPS. In Section \ref{sec:disc}, we examine the use cases for our comprehensive DLPS. Where available, redshifts were used to correct the duration to the rest frame as well as to determine the luminosity distance, assuming a cosmology with H$_0$ = 69.6 km s$^{-1}$ Mpc$^{-1}$, $\Omega_M = 0.286$, and $\Omega_\Lambda = 0.714$.

\section{Data} \label{sec:data}
We assembled X-ray data from a variety of sources. For ease and readability, we include lists of events, as well as their classifications, coordinates, distances, and the relevant literature in Appendix \ref{appA}. Light curve data used in this paper are available on GitHub\footnote{\dataset[https://github.com/avapolzin/X-rayLCs]{https://github.com/avapolzin/X-rayLCs}; we also include some plotting and preliminary light curve classification helper scripts in this repository.}. In Sections \ref{subsec:GRB}-\ref{subsec:ULX+XRB}, we briefly define each (sub)class of transient/variable, describe their characteristic timescale and luminosity as inferred from their position in the DLPS, and detail the provenance of the light curve data used in this work (summarized in Table \ref{tab:class}).

For transient classes -- i.e., events that cannot repeat in the same astrophysical object -- the light curves are from pointed observations acquired after the detection/identification of the transient. As a result, only a subset of the observations capture the peak X-ray luminosity. The situation is somewhat different for variable light curves, some of which include data taken while monitoring the source. We expect that only a fraction of these light curves come from serendipitous detections due to the small field-of-view of the available X-ray instruments.

\begin{figure*}[ht!]
\epsscale{1.2}
\plotone{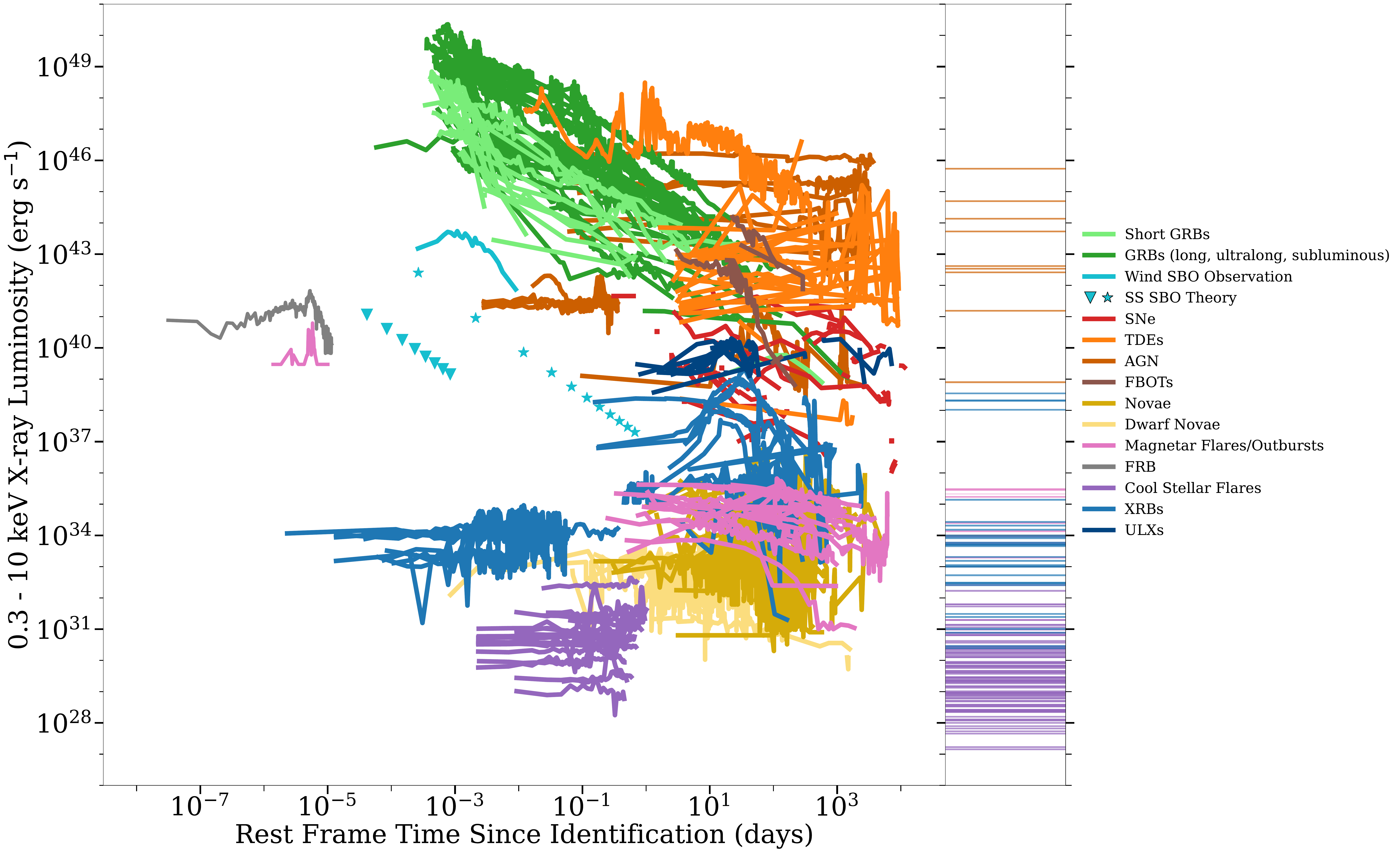}
\caption{
X-ray phase space of transients and variable phenomena, including gamma-ray burst (GRB) afterglows, supernovae (SNe), supernova shock breakouts (SBOs), tidal disruption events (TDEs) and active galactic nuclei (AGN), fast blue optical transients (FBOTs), cataclysmic variables, magnetar flares/outbursts and fast radio bursts, cool stellar flares, X-ray binary outbursts, and ultraluminous X-ray sources. \emph{Main Panel}: X-ray luminosity evolution with rest-frame time since identification. Theoretical SBO peak L$_x$-duration points are shown with different symbols corresponding to the model's input parameters; see Section \ref{subsec:SBO} for details. \emph{Right Side Panel:} To offer a sense of their persistent behavior, the quiescent/pre-flare luminosities of the included variable classes (AGN, magnetar flares/outbursts, cool stellar flares, X-ray binaries, and ultraluminous X-ray sources) are shown as horizontal bars. \label{fig:fullDLPS}}
\end{figure*}

\begin{figure*}[ht!]
\epsscale{1.2}
\plotone{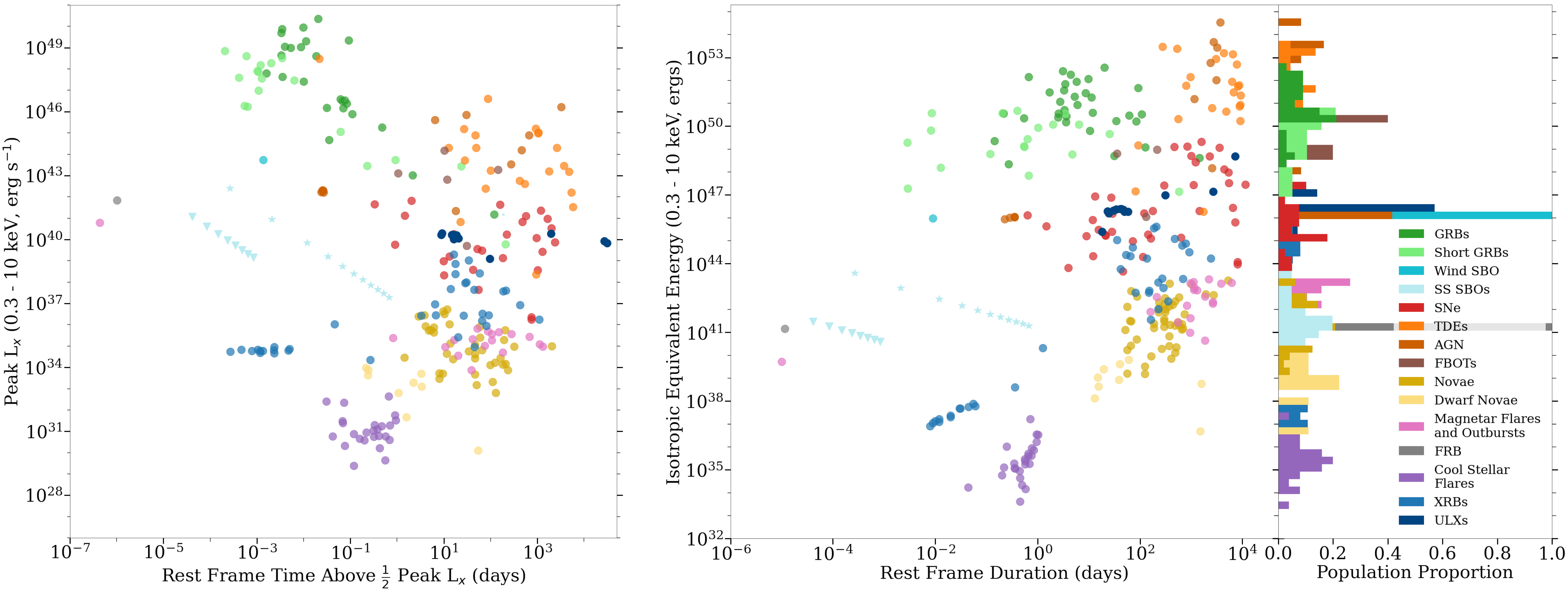}
\caption{\emph{Left Panel:} The peak X-ray luminosity vs. time above half-maximum light. \emph{Right Panel:} At left, the overall energy released during the event vs. the duration of the transient event, and at right, the distribution of isotropic equivalent energies released for each class of transient/variable. As in Figure \ref{fig:fullDLPS}, theoretical SBO peak L$_x$-duration point markers correspond to different input parameters in the model. Points are colored according to the class of transient to which they belong; we use the same color coding as in Figure \ref{fig:fullDLPS} and the histogram at the far right.}\label{fig:fullDLPS_Lpk}
\end{figure*}

These data are used in the observational DLPS (Figure \ref{fig:fullDLPS}), where they show the luminosity evolution of observed light curves with time. These same data are used to calculate important X-ray quantities like peak $L_x$ and isotropic equivalent energy, shown in Figure \ref{fig:fullDLPS_Lpk}, where clustering by class remains apparent. These summary properties provide a means to characterize the light curves regardless of which stage of outburst they cover.

\begin{deluxetable*}{lcc}
\tablenum{1}
\tablecaption{Summary of the classes and subclasses (if any) of transients included in this work. \label{tab:class}}
\tablewidth{\linewidth}
\rotate
\tablehead{Class, $N$ & Subclass, $N$ & Observatories}
\startdata
Gamma-ray Bursts (GRBs), 52 & & \textit{Swift}, \textit{BeppoSAX}, \textit{Chandra}, \textit{XMM-Newton}  \\
  & Short GRBs (sGBRs), 19 & \textit{Swift}, \textit{Chandra} \\
  & Long GRBs (lGRBs), 25 & \textit{Swift} \\
  & Ultralong GRBs, 2 & \textit{Swift} \\
  & Subluminous GRBs, 6 & \textit{Swift}, \textit{BeppoSAX}, \textit{Chandra}, \textit{XMM-Newton} \\
 Shock Breakouts (SBOs), 1 (+6) \tablenotemark{a} & Wind SBO, 1 & \textit{Swift} \\
  & (Stellar Surface SBOs, 6) & (\textit{Swift}, \textit{BeppoSAX}, \textit{Chandra}, \textit{XMM-Newton}) \\
 Supernovae (SNe), 35  & & \textit{Swift}, \textit{BeppoSAX}, \textit{Chandra}, \textit{XMM-Newton}, \textit{ASCA}, \textit{ROSAT} \\
  & Type I Core-Collapse, 9 & \textit{Swift}, \textit{Chandra}, \textit{XMM-Newton}, \textit{ASCA} \\
  & Type II Core-Collapse, 13 & \textit{Swift}, \textit{Chandra}, \textit{XMM-Newton}\\
  & Interacting SNe, 9 & \textit{Swift}, \textit{BeppoSAX}, \textit{Chandra}, \textit{XMM-Newton}, \textit{ASCA}, \textit{ROSAT} \\
  & Superluminous SNe (SLSNe), 2 & \textit{Swift}, \textit{Chandra} \\
  & Ca-rich SNe, 2 & \textit{Swift}, \textit{Chandra} \\
Tidal Disruption Events (TDEs), 19 & & \textit{Swift}, \textit{Chandra}, \textit{XMM-Newton}, \textit{ROSAT} \\
  & Thermal TDEs, 16 & \textit{Swift}, \textit{Chandra}, \textit{XMM-Newton}, \textit{ROSAT}\\
  & Non-thermal TDEs, 3 & \textit{Swift}, \textit{Chandra}, \textit{XMM-Newton}, \textit{ROSAT}\\
Active Galactic Nuclei (AGN), 8 & & \textit{Chandra}, \textit{XMM-Newton}\\
Fast Blue Optical Transients (FBOTs), 5 & & \textit{Swift}, \textit{Chandra}, \textit{XMM-Newton}, \textit{eROSITA}\\
Cataclysmic Variables (CVs), 41 & & \textit{Swift}, \textit{BeppoSAX}, \textit{XMM-Newton}, \textit{ASCA}, \textit{ROSAT}, \textit{RXTE}\\
  & Novae, 38 & \textit{Swift}, \textit{BeppoSAX}, \textit{XMM-Newton}, \textit{ASCA}, \textit{ROSAT}, \textit{RXTE}\\
  & Dwarf Novae, 3 & \textit{Swift}, \textit{RXTE}\\
Magnetar Flares/Outbursts, 15 & & \textit{Swift}, \textit{BeppoSAX}, \textit{Chandra} \textit{XMM-Newton}, \textit{ASCA}, \textit{ROSAT}, \textit{RXTE}, \textit{MAXI}\\
  & Outburst, 14 & \textit{Swift}, \textit{BeppoSAX}, \textit{Chandra} \textit{XMM-Newton}, \textit{ASCA}, \textit{ROSAT}, \textit{RXTE}\\
  & Intermediate Flare/Short Burst, 1 & \textit{Swift}, \textit{MAXI} \\
Fast Radio Bursts (FRBs), 1 & & \textit{Insight-HXMT}\\
Cool Stellar Flares, 18 & & \textit{XMM-Newton}\\
X-ray Binary Outbursts (XRBs), 17 & & \textit{Swift}, \textit{Chandra}, \textit{XMM-Newton}, \textit{RXTE}, \textit{eROSITA}, \textit{NICER}\\
  & Low Mass XRBs (LMXRBs), 4 & \textit{Swift}, \textit{Chandra}, \textit{XMM-Newton}, \textit{RXTE}\\
  & High Mass XRBs (HMXRBs), 13 & \textit{Swift}, \textit{XMM-Newton}, \textit{eROSITA}, \textit{NICER}\\
Ultraluminous X-ray Sources (ULXs), 4 & & \textit{Swift}, \textit{Chandra}, \textit{XMM-Newton}, \textit{ROSAT}\\
\enddata
\tablecomments{$N$ indicates the number of included objects. We also list the observatories that were used in the creation of the light curves (including upper-limits). Observatories used include: Swift \citep{2005SSRv..120..165B}, BeppoSAX \citep{1997AAS..122..299B}, Chandra \citep{2000SPIE.4012....2W}, the X-ray Multi-Mirror Mission \citep[XMM-Newton;][]{2001AA...365L...1J}, the Advanced Satellite for Cosmology and Astrophysics \citep[ASCA;][]{1994PASJ...46L..37T}, the Roentgen Satellite \citep[ROSAT;][]{trumper_1990}, the Spectrum Roentgen Gamma (SRG) extended Roentgen Survey with an Imaging Telescope Array \citep[eROSITA;][]{2021AA...647A...1P}, the Rossi X-ray Timing Explorer \citep[RXTE;][]{1996SPIE.2808...59J}, the Monitor of All-Sky X-ray Image \citep[MAXI;][]{2009PASJ...61..999M}, the Hard X-ray Modulation Telescope \citep[Insight-HXMT;][]{2020SCPMA..6349502Z}, and the Neutron Star Interior Composition Explorer \citep[NICER;][]{2014SPIE.9144E..20A}.}
\tablenotetext{a}{We represent the six \textit{candidate} stellar surface shock breakouts, which overlap entirely with the population of subluminous GRBs, in parentheses for completeness.}
\end{deluxetable*}

\begin{figure*}[ht!]
\epsscale{1.09}
\plotone{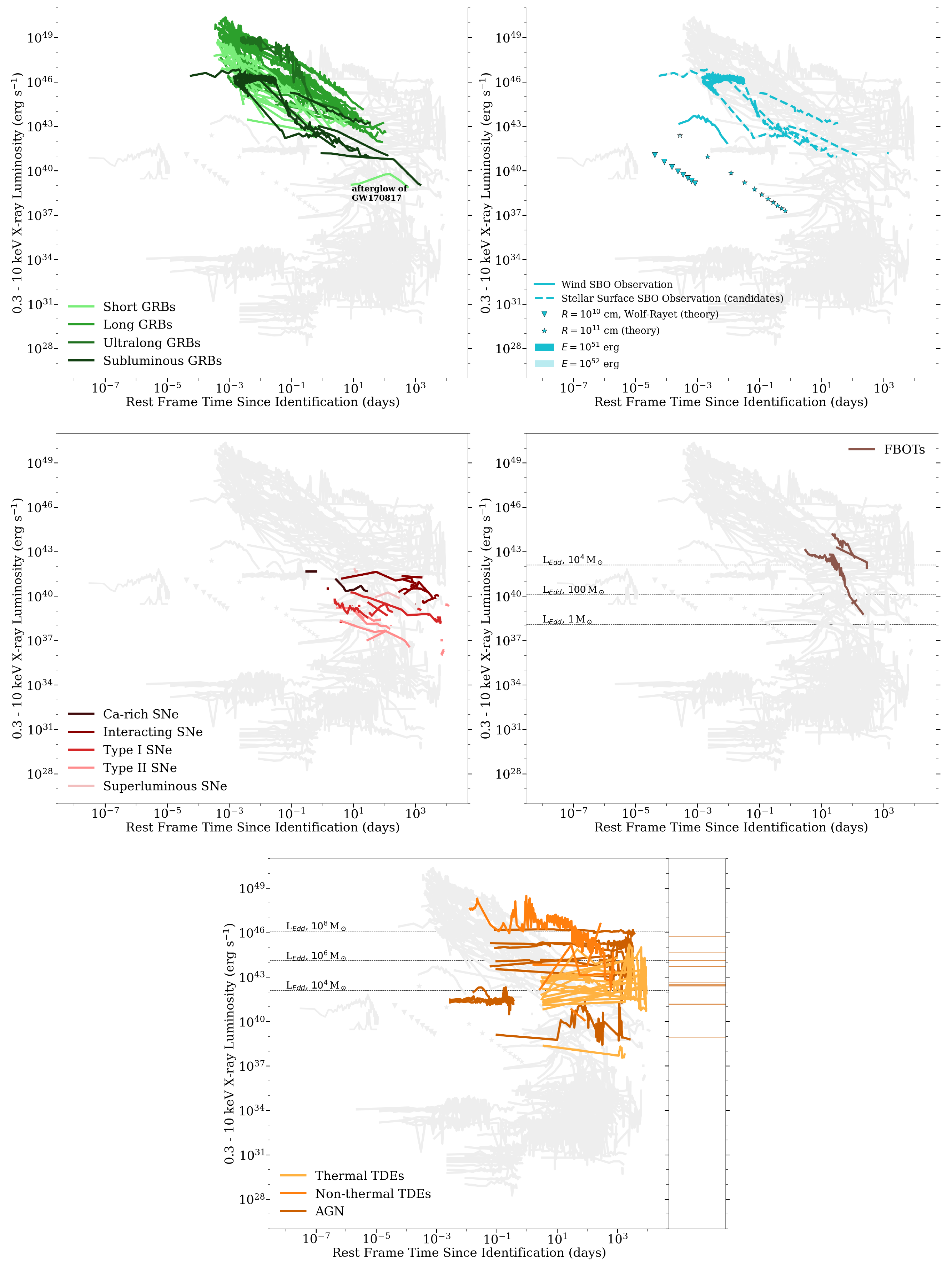}
\caption{X-ray phase space of extragalactic transients, including GRBs, SBOs, SNe, TDEs, AGN, and FBOTs, with all other classes of transient from this work underplotted in gray. Peak luminosity vs. duration are shown as points for modeled SBOs \citep{2012ApJ...747...88N} with input progenitor radius and breakout energy indicated in the legend by $R$ and $E$ respectively. We underplot Eddington luminosities (as horizontal dashed lines) for some potentially relevant BH progenitor masses for both FBOTs and TDEs/AGN \citep{2022MNRAS.515.1146R, 2023arXiv230306523Y}. The FBOT X-ray counterparts occupy a luminosity range that is intermediate between normal SNe (shades of red) and GRBs (shades of green). At right in the TDE/AGN subplot, we show pre-flare AGN luminosities for all included AGNs. Included events are listed in Tables \ref{tab:GRB} through \ref{tab:FBOT}. \label{fig:GRBDLPS} \label{fig:SBODLPS} \label{fig:SNeDLPS} \label{fig:TDEDLPS} \label{fig:FBOTDLPS}}
\end{figure*}

\subsection{Gamma-ray Burst (GRB) Afterglows } \label{subsec:GRB}
Gamma-ray bursts (GRBs, Table \ref{tab:GRB}) are burst of $\gamma$-rays associated with either the collapse of a massive star (GRBs with a duration of the $\gamma$-ray emission $T_{90}>2$ s) or the merger of compact objects (i.e. neutron stars and black holes).
All GRB X-ray afterglow data shown in Figure \ref{fig:GRBDLPS} were collected via the
UK \emph{Swift} Science Data Centre\footnote{\href{https://www.swift.ac.uk}{https://www.swift.ac.uk}} \citep{2007A-A...469..379E, 2009MNRAS.397.1177E}, with the notable exceptions of the pre-\emph{Swift} era subluminous GRBs, GRB980425A \citep{2000ApJ...536..778P, 2004ApJ...608..872K} and GRB031203A \citep{2004Natur.430..646S, 2004ApJ...605L.101W}. We include long GRBs (lGRBs), short  GRBs (sGRBs), ultralong GRBs, and subluminous GRBs for $z \le 1$ in our plotted population where redshift information is available\footnote{\href{http://www.mpe.mpg.de/~jcg/grbgen.html}{http://www.mpe.mpg.de/$\sim$jcg/grbgen.html}}. We also include the X-ray afterglow counterpart of the neutron-star merger event GW170817, for which gravitational-wave emission was detected 
\citep[e.g.,][]{2017ApJ...848L..12A, 2019ApJ...886L..17H, 2020RNAAS...4...68H, 2020PhR...886....1N, 2021ARAA..59..155M}.  We excluded GRBs without well-constrained redshifts as we are interested in luminosity vs. intrinsic duration (rather than fluence vs. observed duration).

Our sample is complete for subluminous and ultralong GRBs. We include all but one of the long GRBs within $z = 1$ with X-ray observations from the end of December 2014 through 2019, and all but one of the X-ray observations of short GRBs within $z = 1$ from 2005 through 2017.

Differentiation of the subclasses of GRBs was informed by the $T_{90}$ parameter (i.e. the time interval over which 90\% of the $\gamma$-ray emission is observed). Short GRBs typically have $T_{90} < $2\,s, long GRBs fall within the 2 -- $10^3$\,s range \citep[][]{1993ApJ...413L.101K}, and ultralong GRBs have $T_{90}$ between $10^3$ -- $10^4$\,s \citep[][]{2014ApJ...781...13L}. We note that some subluminous GRBs, while having a duration similar to that of long or ultralong GRBs, might actually represent physically distinct phenomena (e.g. supernova shock breakouts see Section \ref{subsec:SBO}) with $L_x \lesssim 10^{47}$ erg s$^{-1}$ \citep{2003AIPC..686...74N}.

\subsection{Explosion Shock Breakouts} 

Shock breakouts (SBOs, Table \ref{tab:SBO}) are the emergence of the first (observable) photons from a stellar explosion. A SBO occurs as the shock goes through the star and reaches an optical depth of $\tau \sim c/v_{\rm{shock}}$ within the star or at the stellar surface or in the stellar wind. Short-duration energetic emission is observable in the X-rays if the shock breaks out from a compact progenitor \citep[][]{2010ApJ...725..904N}. SBOs are short duration when their emission peaks in the X-rays, and there is only one broadly accepted observation (\citealt{2008Natur.453..469S}, see however \citealt{2008Sci...321.1185M} for a different interpretation), which was a serendipitous detection from a normal type Ib supernova, SN\,2008D. While searches of archival data yield potential SBO candidates \citep[e.g.][]{2020ApJ...896...39A, 2020ApJ...898...37N}, wide-field X-ray instruments are vital for growing the sample of SBO observations.
\label{subsec:SBO}
We note that later analysis of the prompt X-ray signal at the location of SN\,2008D showed what is thought to be a breakout from the stellar wind \citep[][]{2011MNRAS.414.1715B, 2014ApJ...788L..14S}. We tentatively include subluminous GRBs as candidate stellar surface breakouts associated with energetic type Ic-BL supernovae in Figure \ref{fig:SBODLPS}. Subluminous GRBs are considered candidate stellar surface breakouts by \citet{2012ApJ...747...88N}, and this possibility is also addressed by \citet{2006Natur.442.1008C}, \citet{2015ApJ...807..172N}, and \citet{2021MNRAS.508.5766I}.

EXMM 023135.0-603743 \citep{2020ApJ...896...39A, 2020ApJ...898...37N}, on the other hand, is not included as a candidate in the DLPS due to the uncertain nature of its progenitor. In addition to the possibility that it is a shock breakout from a core-collapse supernova, both \citet{2020ApJ...896...39A} and \citet{2020ApJ...898...37N} discuss alternative physical scenarios that could give rise to the observed X-ray transient.

In order to better populate the X-ray phase space (Figure \ref{fig:SBODLPS}), we supplement the proposed stellar surface SBO light curves (from subluminous GRBs) with results from theoretical calculations by \citet{2012ApJ...747...88N}. These authors show that:

\begin{equation} \label{eq:Ebo}
    E_{bo} \approx 6\times10^{46}\,E_{53}^{2.3}M_{ej,5}^{-1.65}R_5^{0.7} \; \textnormal{erg}
\end{equation}
\begin{equation}
    T_{bo} \approx 700\,E_{53}^{1.7}M_{ej,5}^{-1.2}R_5^{-0.95} \; \textnormal{keV} \label{eq:T}
\end{equation}
\begin{equation} \label{eq:tbo}
    t_{bo}^{obs} \approx 0.06\, E_{53}^{-3.4}M_{ej,5}^{2.5}R_5^{2.9} \; \textnormal{s}
\end{equation}
\begin{equation} \label{eq:Lbo}
    L_{bo} \approx 4\times10^{47}\,E_{53}^{5.1}M_{ej,5}^{-3.65}R_5^{-1.85} \; \textnormal{erg  s}^{-1}
\end{equation}
Where $E_{bo}$, $T_{bo}$, $t_{bo}^{obs},$ and $L_{bo}$ refer to the breakout energy, temperature, observed time (duration), and luminosity respectively. $E_{53}$ is energy in terms of $10^{53}$ erg, $M_{ej, 5}$ is the ejecta mass in terms of $5M_{\odot}$, and $R_5$ is the stellar radius in terms of $5R_{\odot}$.

We use a grid of energy values (between $10^{51}$ and $10^{52}$ erg), ejecta mass (between 1 and 10$M_\odot$), and stellar radius ($10^{10}$ and $10^{11}$ \,cm, which span the properties of Wolf-Rayet-like stars) to compute $E_{bo}$, $T_{bo}$, $t_{bo}$, and $L_{bo}$ from Equations \ref{eq:Ebo} - \ref{eq:Lbo}.
Though a red supergiant with a breakout energy of $10^{52}$ erg is a less likely physical scenario, we include it anyway to account for the full range of possible progenitors that give rise to SBOs within the phase space. Similarly, in order to populate the phase space with potential durations vs. peak luminosities, we limit our plotted sample to those with temperatures (from Equation \ref{eq:T}) in the range 0.1-20 keV as representative of the SBOs that will  have some X-ray luminosity component in the 0.3-10 keV range of interest. In the upper-right panel of Figure \ref{fig:SBODLPS}, each individual point represents the peak luminosity and duration of a single theoretical stellar surface SBO event.

Shock breakouts from the stellar wind like in SN2008D evolve on timescales ranging from seconds to minutes with $L_x \sim 10^{42} - 10^{44}$ erg s$^{-1}$, making it challenging to observe them without wide field of view X-ray instruments facilitating serendipitous detection. Candidate stellar surface SBOs (as potentially in subluminous GRBs) range in luminosity from $\sim 10^{41} - 10^{47}$ erg s$^{-1}$, varying on timescales of $\sim 10^{-4} - 10^{-1}$ days. Our modeled stellar surface SBOs \citep{2012ApJ...747...88N} are somewhat less luminous and shorter-lived, with $L_x \sim$ a few $\times\, 10^{36} - 10^{43}$ erg s$^{-1}$ and $t \sim$ several $\times\, 10^{-5}$ days to $\sim 1$ day.
 
\subsection{Supernovae} \label{subsec:SN}
Supernova (SN, Table \ref{tab:SNe}) shocks that propagate in the explosion's environment are well-known particle accelerators and well-known sources of X-ray emission as the shocks decelerate and the particles cool down (e.g., \citealt{2017hsn..book..875C} for a recent review).
We collected X-ray data for supernovae from a variety of sources (see Appendix \ref{appA} for details). 
Because of the rather limited sample of existing observations, we include all available ($z \le 1$) X-ray light curves in bands with lower energy limits between 0.2 and 0.5 keV and upper energy limits between 8 and 12 keV, which are then k-corrected to the observed 0.3-10 keV energy band assuming a spectrum $F_{\nu} \propto \nu^{-\beta}$ with a spectral index $\beta = 1$ (equivalent to a photon index $\Gamma = 2$), which is consistent with observed spectral properties of SNe in the X-rays \citep[e.g.,][]{2011arXiv1109.0981L}. Even at the most extreme ends of our allowed input energy limits, using $\Gamma = 1$ or $\Gamma = 3$ instead represents a difference of less than a factor of two in luminosity. These k-corrected data are shown in Figure \ref{fig:SNeDLPS}.

We divide the SNe into three subclasses based on their underlying physical properties: \textit{Type I core-collapse} to be comprised of Type Ib, Ic, Ib/c, Ic/pec, and IIb SNe; \textit{Type II core-collapse} to be comprised of Type II, IIP, IIL, and IIpec SNe; and \textit{Interacting SNe} (i.e. SNe with signatures of CSM interaction in their optical spectra) to be comprised of Type IIn, Ibn, and Ia-CSM SNe. Additionally, we designate (optically) \textit{superluminous SNe} (SLSNe) and \textit{Ca-rich SNe} separately as the two subclasses of SNe for which X-ray emission has been most recently found.

Unlike GRBs, SNe are generally not monitored in the X-rays, in part because they are intrinsically much fainter in the X-rays than GRBs. As a result, they have relatively sparse observations. We include what (non-upper limit) detections are available in the DLPS. We are complete with respect to published X-ray light curves of SNe through the end of 2012, and we have tried to be complete for all non-ordinary SNe (Ca-rich, superluminous, Ia-CSM) with X-ray detections by the time of submission.

Within the DLPS, SNe evolve on timescales ranging $10^{-1} - 10^{4}$ days. Ca-rich and interacting SNe have luminosities $\sim 10^{39} - 10^{42}$ erg s$^{-1}$ with Ca-rich supernovae evolving on timescales between $0.1$ and $\sim$ hundreds of days and interacting SNe evolving on timescales between 1 day and thousands of days. In general, optically superluminous supernovae are less luminous in the soft X-rays with typical $L_x \sim $ several $\times 10^{40} - \mathrm{several} \times 10^{41}$ erg s$^{-1}$. Type I core-collapse SNe are slightly less luminous with $L_x \sim 10^{38} - 10^{40}$ erg s$^{-1}$, and Type II core-collapse supernovae are the least luminous with most light curves spanning $L_x \sim$ several $\times 10^{35} - 10^{39}$ erg s$^{-1}$. Further discussion of the differences in the observed X-ray light curves of different classes of SNe can be found in \citet{2012MNRAS.419.1515D} (see \citealt{2021ApJ...908...75B} for a similar discussion in the radio).

\subsection{Tidal Disruption Events and Active Galactic Nuclei} \label{subsec:TDE}
Tidal disruption events (TDEs, Table \ref{tab:TDE}) occur when a star passes close enough to a black hole that stellar material is accreted, resulting in high energy electromagnetic emission from that accretion \citep[][]{1982Natur.296..211C, 1983AA...121...97C}.

We include both TDEs with thermal X-ray emission and non-thermal X-ray emission in Figure \ref{fig:TDEDLPS}, using \citet{2015JHEAp...7..148K} and \citet{2017ApJ...838..149A} to inform our selection of ($z \le 1$) TDE candidates, showing only \textit{X-ray TDEs} and ``\textit{Likely X-ray TDEs}'' from the latter. Our sample of tidal disruption events is complete (and robust – for merging multiple catalogs) until 2017.

TDEs with non-thermal X-ray emission (from hereon, non-thermal TDEs) belong to a subset of $\sim 10 \%$ the TDE population that showed evidence for highly collimated ejecta in the form of relativistic jets \citep[][]{2020SSRv..216...81A}. There is no evidence for collimation of the thermal X-ray emission which implies that TDEs with thermal X-rays (from hereon, thermal TDEs) are easier to detect. 
Because there might be similarities between the flare mechanisms of TDEs and active galactic nuclei (AGN) and the distinction between the two classes can be observationally challenging, we opt to show them both in the bottom panel of Figure \ref{fig:TDEDLPS}. In the interest of examining only flaring/outbursting behavior, we include long-term variability from AGN, while we exclude changing-look AGN, which exhibit more persistent variability. We convert the sample of light curves \citep[][]{2018ApJ...852...37A} to our 0.3-10 keV energy band, assuming $\Gamma = 1.8$ \citep[][]{2006AA...451..457T}. We note that, though we are far from showing \textit{all} AGN light curves in this energy band, we aim to show a representative sample which demonstrates the difficulty in separating TDEs and AGN from light curves alone (for additional AGN/blazar light curves, see e.g., \citealt{2019AA...631A.116G}).

The Quasi-Periodic Eruptions (QPEs) from GSN 069 \citep[][]{2019Natur.573..381M} occupy a slightly different (similar luminosity, shorter duration) part of the phase space, with $L_x\sim$10$^{41} - 10^{42}$ erg s$^{-1}$ and $\sim$10$^{-3} - 10^{-1}$ days (AGN have X-ray luminosities between a few $\times$10$^{38}$ and $\gtrsim$10$^{46}$ erg s$^{-1}$ and vary on timescales of $10^{-1}$ to $10^{3}$ days; here the low luminosity end of the range is set by NGC 4395, see \citealt{2018ApJ...852...37A}). The physical mechanism that drives QPEs is not fully understood, so we include an example for consistency in looking at AGN outbursting activity, though they may be associated with the same mechanism as changing-look AGN.

While the archetypal non-thermal TDE Swift 1644+57 was initially mistaken for a long GRB, Figure \ref{fig:TDEDLPS} shows that TDEs are clearly distinguished from GRBs for their luminous (non-thermal TDEs have luminosities between $10^{42}$ and 10$^{49}$ erg s$^{-1}$, while thermal TDEs are somewhat less luminous with $L_x \sim$10$^{37} - 10^{45}$ erg s$^{-1}$)\, \emph{and persistent} X-ray emission lasting hundreds of days.

\subsection{Fast Blue Optical Transients}
Fast blue optical transients (FBOTs, Table \ref{tab:FBOT}) are a new class of transient astronomical event, only recently recognized in observations and the literature \citep[e.g.,][]{2014ApJ...794...23D, 2016ApJ...819...35A, 2016ApJ...819....5T, 2018MNRAS.481..894P, 2021arXiv210508811H}. In the optical bands these transients are characterized by short rise times (evolution on the timescale of days) and can reach high luminosities ($L \gtrsim 10^{44}$ erg s$^{-1}$).  We include the five known  (as of October 2022) X-ray instances of this class -- CSS161010 \citep{2020ApJ...895L..23C}, AT2018cow \citep{2018MNRAS.480L.146R, Margutti_2019, 2019MNRAS.484L...7R}, AT2020xnd \citep{2022ApJ...926..112B, 2022ApJ...932..116H}, AT2020mrf \citep[][]{2022ApJ...934..104Y}, and AT2022tsd \citep[][]{2022TNSAN.207....1S, 2022TNSAN.218....1M} -- in our phase space plot, Figure \ref{fig:FBOTDLPS}. Until now, only the most luminous optical FBOTs (collectively referred to as Luminous FBOTs -- LFBOTs) have exhibited detectable X-ray counterparts.

\subsection{Cataclysmic Variables}

Cataclysmic variables (CVs, Table \ref{tab:CVs}) are binary systems undergoing mass transfer in which a white dwarf accretes material from a low-mass main sequence companion. The conditions of that mass transfer define the characteristics of the CV outburst. Here we look exclusively at the two classes of CVs that exhibit bursting behavior\footnote{We do not include light curves from non-outbursting CVs, like polars, which exhibit low amplitude flickering and flaring due to their stronger magnetic fields \citep[e.g.,][]{1989MNRAS.238..697A}.} -- (classical and recurrent) novae and dwarf novae.

\begin{figure*}[ht!]
\epsscale{1.2}
\plotone{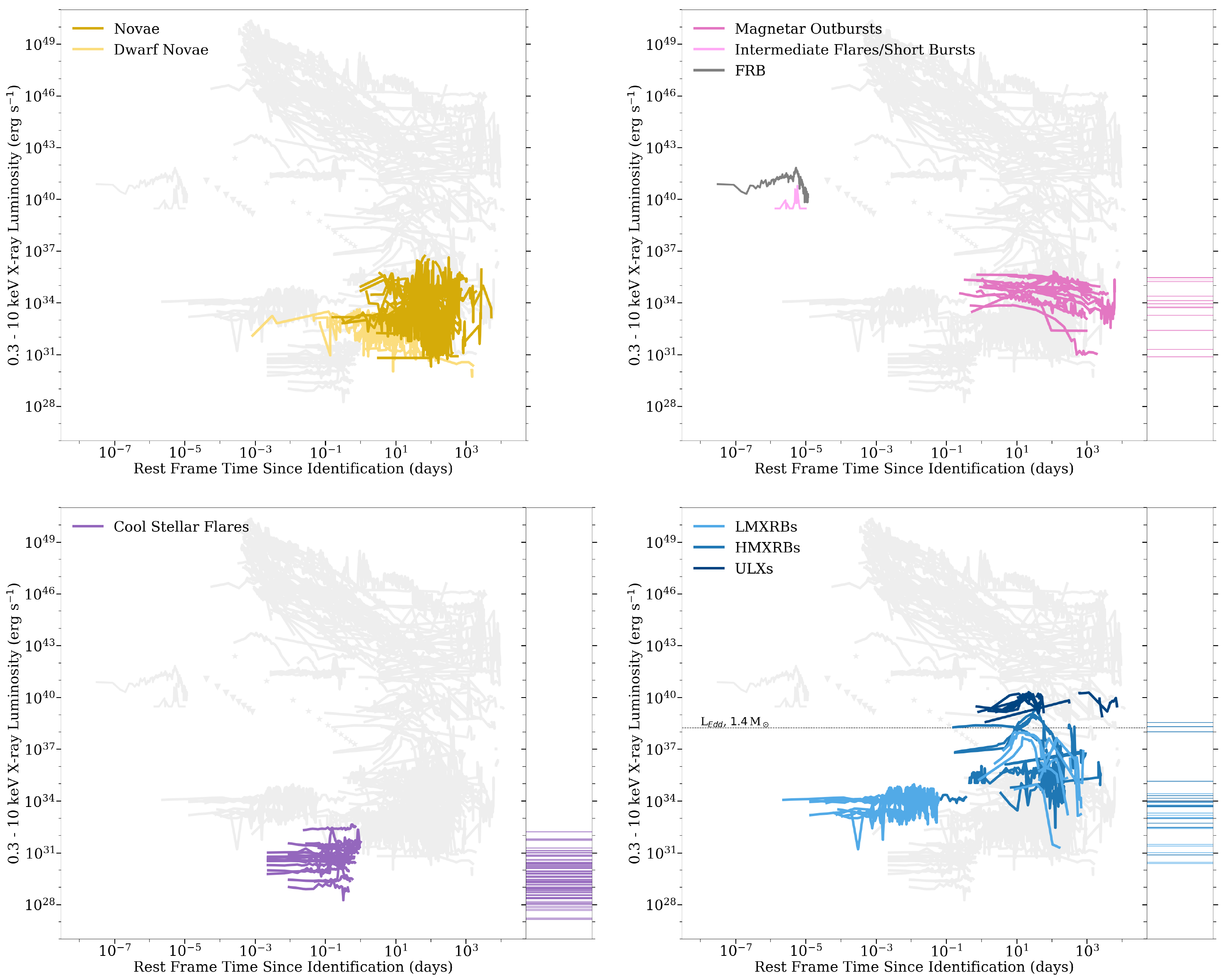}
\caption{X-ray phase space of Galactic (CVs, magnetar flares/outbursts, and cool stellar flares) and Galactic/extragalactic (XRBs, ULXs, and FRBs) transients and variables, including those classes of event with some signals which have been observed to originate within the Galaxy, such as CVs (novae and dwarf novae), magnetar flares and outbursts, fast radio bursts (specifically the Galactic FRB200428), cool stellar flares, XRBs, and ULXs. We underplot the Eddington luminosity for a h$1.4\, M_\odot$ progenitor for additional context in the XRBs/ULX panel. With the exception of the CV subplot, at right, we show quiescent luminosities for each class of object. Included events are listed in Tables \ref{tab:novae} through \ref{tab:XRB-ULX}. 
\label{fig:NovaeDLPS} \label{fig:MagnetarDLPS} \label{fig:MdwarfDLPS} \label{fig:ULXXRBDLPS}}
\end{figure*}

A classical/recurrent nova outburst occurs when the accreted material causes thermonuclear runaway on the surface of the white dwarf, resulting in highly energetic ejection of material from the stellar surface. All X-ray data of these novae are from \citet{2008ApJ...677.1248M} and \citet{2020AdSpR..66.1169P}; we include all classical/recurrent novae detected between 2006 and 2017. In the instances where we have both an upper- and lower-limit luminosity for various novae in \citet{2008ApJ...677.1248M}, we utilized both, to return a \textit{lower-} and \textit{upper-}limit light curve, offering us a greater sense of where novae can, and do, exist in the X-ray phase space (Figure \ref{fig:NovaeDLPS}).
Where k-corrections are necessary to shift data into the (observer frame) 0.3-10 keV energy band, we adopt a thermal bremsstrahlung spectral model with $kT = 5$ keV following \citet{2008ApJ...677.1248M}. 

The X-ray emission in dwarf novae stems from the inner accretion flow region around the white dwarf. During the outburst, the mass transfer rate through this inner region increases. As a result it is expected that the X-ray emission ($\geq2$ keV) will briefly increase, but then be suppressed as the optical depth of this region increases. This behavior can be seen in multi-wavelength light curves of the dwarf nova SS Cygni \citep[e.g.,][]{2003MNRAS.345...49W, 2016MNRAS.460.3720R}. Note that there are a number of unanswered questions about this model \citep[see][]{2017PASP..129f2001M}, and no other dwarf novae show this exact behavior \citep[e.g.,][]{2011PASP..123.1054F, 2017PASP..129f2001M}.  As we are interested in the DLPS of systems that show X-ray brightenings, we limit our sample to those dwarf novae that show X-ray brightenings during optical outburst (see Table \ref{tab:CVs}). The three dwarf novae included here are those with good temporal coverage and multi-wavelength data that supports enhancement in the X-rays. There are $\sim8$ other DNe that show this same brightening, but each has only $\sim1$ detection in outburst.

CV outbursts are fairly low luminosity in the $L_x \sim 10^{28} - 10^{36}$ erg s$^{-1}$ range, with dwarf novae only reaching a peak $L_x \sim$10$^{34}$ erg s$^{-1}$ and classical/recurrent novae spanning that entire range. They evolve on timescales ranging from seconds to years.

\subsection{Magnetar Flares/Outbursts and Fast Radio Bursts} \label{subsec:magnetar}

Magnetar flares/outbursts (Table \ref{tab:magnetars}), driven by perturbations in the strong magnetic field of the magnetar, come in three broad flavors: giant flares (to date, only observed in the hard X-rays and gamma rays), outbursts (characterized by a decay on the scale of days), and intermediate flares/short bursts (lasting milliseconds to tens of seconds).
At gamma-ray energies, the three observed giant flares started with a short (0.1--0.2\,s) flash with luminosity from $\approx 10^{44}$ to $10^{46}$ erg s$^{-1}$, which was followed by a tail lasting a few hundreds of seconds and modulated at the pulsar spin period. In all three events, the total energy of the tail was $\approx 10^{44}$ erg (e.g. \citealt{2021ASSL..461...97E}). While it is likely that a comparable amount of energy was emitted in the soft X-ray band (see \citealt{2013ApJ...775L..34R}), we lack reliable measurements of their properties in that band.

Though intermediate flares and short bursts are often referred to separately, \citet{2008ApJ...685.1114I} suggest that these events actually occur along a continuum of spectral properties (though not a continuum in duration or fluence). Making an arbitrary cut, where intermediate flares persist longer and are brighter while short bursts are lower energy and shorter duration, is not based on intrinsically different physics. For the purposes of simplicity in our sub-classifications, we consider intermediate flares and short bursts to be a single population, characterized by $L_x \sim$10$^{39} - 10^{41}$ erg s$^{-1}$ and varying on extremely short timescales $\sim$10$^{-6} - 10^{-4}$ days.

We used the Magnetar Outburst Online Catalog \citep{2018MNRAS.474..961C} for most of the \textit{outburst} data ($L_x \sim$10$^{31} - 10^{46}$ erg s$^{-1}$ with variation on timescales $\sim$10$^{-1} - 10^{4}$ days), also including data from \citet{2016ApJ...828L..13R} and \citet{2019A-A...626A..19E} in order to build a \textit{representative} sample. Plotting each light curve from the beginning of the outburst itself, we show each recurrent event from the same progenitor separately. 

In order to elucidate the variable nature of these magnetars, we compare their luminosity in outburst (or during a flare) to their quiescent luminosity; we retrieve these data from \citet{Olausen_2014}\footnote{\href{http://www.physics.mcgill.ca/~pulsar/magnetar/main.html}{http://www.physics.mcgill.ca/$\sim$pulsar/magnetar/main.html}} in the 2-10 keV energy band. We employ a k-correction, to appropriately relate these luminosities to the 0.3-10 keV behavior we have emphasized throughout the X-ray phase space plot, given a power law, black body, or power law + black body model individual to the source from \citet{Olausen_2014}. Where a spectral fit is not offered, we use a generalized multiple component spectrum ($\Gamma \sim 2$, $kT_{low} \sim 0.3$ keV, $kT_{high} \sim 0.6$ keV) in quiescence \citep{Mong_2018}. Further, to ensure a one-to-one comparison of the emission from quiescent magnetars and those actively exhibiting variable behavior, we restrict our quiescent $L_x$ sample to match the magnetars shown in the X-ray phase space plot (Figure \ref{fig:MagnetarDLPS}). 

Due to the extremely fleeting nature of the short bursts and intermediate flares, much of the data comes from serendipitous triggers, many of which occur in the harder X-rays, since the current class of wide-field instruments operate in the hard X-rays/gamma-rays. This accounts, in part (or in whole), for the paucity of observations for these phenomena in the soft X-rays (and so in our phase space plot) relative to the frequency with which they occur.

Fast Radio Bursts (FRBs) are extremely short duration transient events characterized by an intense burst of radio emission (\citealt{2007Sci...318..777L}; or see \citealt{2019AARv..27....4P} and \citealt{2022AARv..30....2P} for reviews). Multiwavelength follow-up has been conducted to detect counterparts in other wavelength regimes, but efforts have been largely unsuccessful \citep[e.g.,][]{2020ApJ...897..146C}. Recently, though, the SGR 1935+2154 outburst on 2020 April 28 has been the subject of discussion as a candidate for an X-ray counterpart to FRB 200428 \citep{2020Natur.587...54T, 2020Natur.587...59B}. Concurrent radio and X-ray emission from this source was detected again in October 2022 \citep[][]{2022ATel15681....1D, 2022ATel15682....1W}.

Because there is evidence linking this event to a magnetar progenitor, we include the FRB light curve \citep[][]{2021NatAs...5..378L} in Figure \ref{fig:MagnetarDLPS}. The coincident X-ray event from SGR 1935+2154 is consistent with the apparent continuum behavior of magnetar outbursts, flares, and bursts across the phase space (with $L_x \sim$10$^{40} - 10^{42}$ erg s$^{-1}$ on timescales $\sim$10$^{-8} - 10^{-5}$ days) and is indicative of the possibility that some FRBs might be the radio counterparts to soft gamma repeaters  (see however \citealt[][]{2021ApJ...923....1P}).

\subsection{Cool Stellar Flares} \label{subsec:M-dwarf}

Cool, low mass stars (Table \ref{tab:mdwarfs}), such as M-dwarfs, can be highly variable, with energetic flares driven by magnetic reconnection events. The intensity of this behavior is also correlated with age, with younger low-mass stars exhibiting more variability.

We place data from \citet{2015A-A...581A..28P} in Figure \ref{fig:MdwarfDLPS} and assume a thermal spectral model with a temperature of $kT=$1.5 keV in order to perform the flux conversion. Dwarf stars included in our sample are K, M, and L types. As in Section \ref{subsec:magnetar}, quiescent luminosities (digitized from \citealt{2015A-A...581A..28P}) are plotted at the right to appropriately contextualize the flares and offer yet more indication of where these flaring stars exist in the X-ray phase space. Cool stellar flares are relatively short duration, with timescales ranging from on the order of hundreds of seconds up to $\sim 1$ day. They are also low luminosity\footnote{We note that all sky survey data has shown intrinsically rare flares up to $L_x \sim 10^{34}$  erg s$^{-1}$\ at slightly higher energies (2 - 20 keV; \citealt{2016PASJ...68...90T}).} events with L$_x \gtrsim 10^{28}$ erg s$^{-1}$ and up to several $\times \; 10^{32}$ erg s$^{-1}$. Quiescent luminosities span the range $\sim 10^{27} - 10^{32}$ erg s$^{-1}$.

\subsection{X-ray Binary Outbursts and Ultraluminous X-ray Sources} \label{subsec:ULX+XRB}

X-ray binaries (XRBs, Table \ref{tab:XRB-ULX}) are stellar binaries where a compact object (neutron star or black hole) is accreting material from its companion, causing energetic outbursts. Ultraluminous X-ray sources (ULXs, characterized by peak L$_x > 10^{39}$ erg s$^{-1}$, independent of the source's underlying mechanism) are frequently associated with super-Eddington XRBs. We also elect to group them here, showing them in the same subplot of Figure \ref{fig:ULXXRBDLPS}. XRBs are further broken out into high mass (HMXRBs, with a companion star of mass $\gtrsim 10$ M$_\odot$; for a review, see \citealt{2011Ap&SS.332....1R}) and low mass (LMXRBs, generally with a $M \lesssim 1.5$ M$_\odot$ companion; for a review, see \citealt{2016ApJS..222...15T}) populations. The former includes BeXRBs, supergiant X-ray binaries (SGXRBs), and supergiant fast X-ray transients (SFXTs), while the latter includes neutron star X-ray binaries (NS-XRBs), black hole X-ray binaries (BH-XRBs), and, though we do not have any in our DLPS sample, very faint X-ray transients (VFXTs; \citealt{2015MNRAS.447.3034H}) or very faint XRBs. Details of the relevant data and their provenance are in Table \ref{tab:XRB-ULX}. 

As with the other variable signals (magnetar outbursts and cool stellar flares, top right and bottom left of Figure \ref{fig:MdwarfDLPS}), we plot the quiescent L$_x$ of ULX and XRB events at the right in Figure \ref{fig:ULXXRBDLPS}. As in Section \ref{subsec:magnetar}, for the purposes of this paper, we define the quiescent luminosity as the lowest recorded X-ray luminosity, opting for simplicity rather than a more stringent definition that might not designate this persistent, non-outburst behavior as quiescence. Though we are not aiming for completeness, choosing instead to use a \textit{representative} sample, the XRB and ULX coverage of the phase space is clear for relatively long timescales ranging from tenths to thousands of days and intermediate luminosities. XRBs exist in roughly the $L_x \sim 10^{32} - 10^{39}$ erg s$^{-1}$ range (with quiescent luminosities between $10^{30}$ and $10^{35}$ erg s$^{-1}$ and outburst $L_x \gtrsim 10^{35}$ erg s$^{-1}$). ULXs have luminosities greater than several $\times \; 10^{38}$ erg s$^{-1}$ and up to $\sim 10^{40}$ erg s$^{-1}$, with quiescent L$_x$ falling between $10^{38}$ and $10^{39}$ erg s$^{-1}$.

In regards to target selection, it is necessary to use a sample of sources with well-measured distances. We note that many of the Galactic HMXRBs have poorly measured distances with high uncertainties \citep[e.g.][]{2019A-A...622A..93B,2022AA...664A..99F}, while their soft X-ray spectra might suffer from strong absorption. However, nearby galaxies of the Small and Large Magellanic Clouds (i.e. SMC and LMC) have well defined distances, low foreground absorption and an abundance of HMXRBs \citep{2016A&A...586A..81H}. Thus a representative sample of HMXRB outbursts was obtained from the SMC and LMC.

\section{Discussion} \label{sec:disc}
\subsection{Unclassified X-ray Sources: A Short Case Study} \label{subsec:unknown}

\begin{figure*}[ht!]
\epsscale{1.2}
\plotone{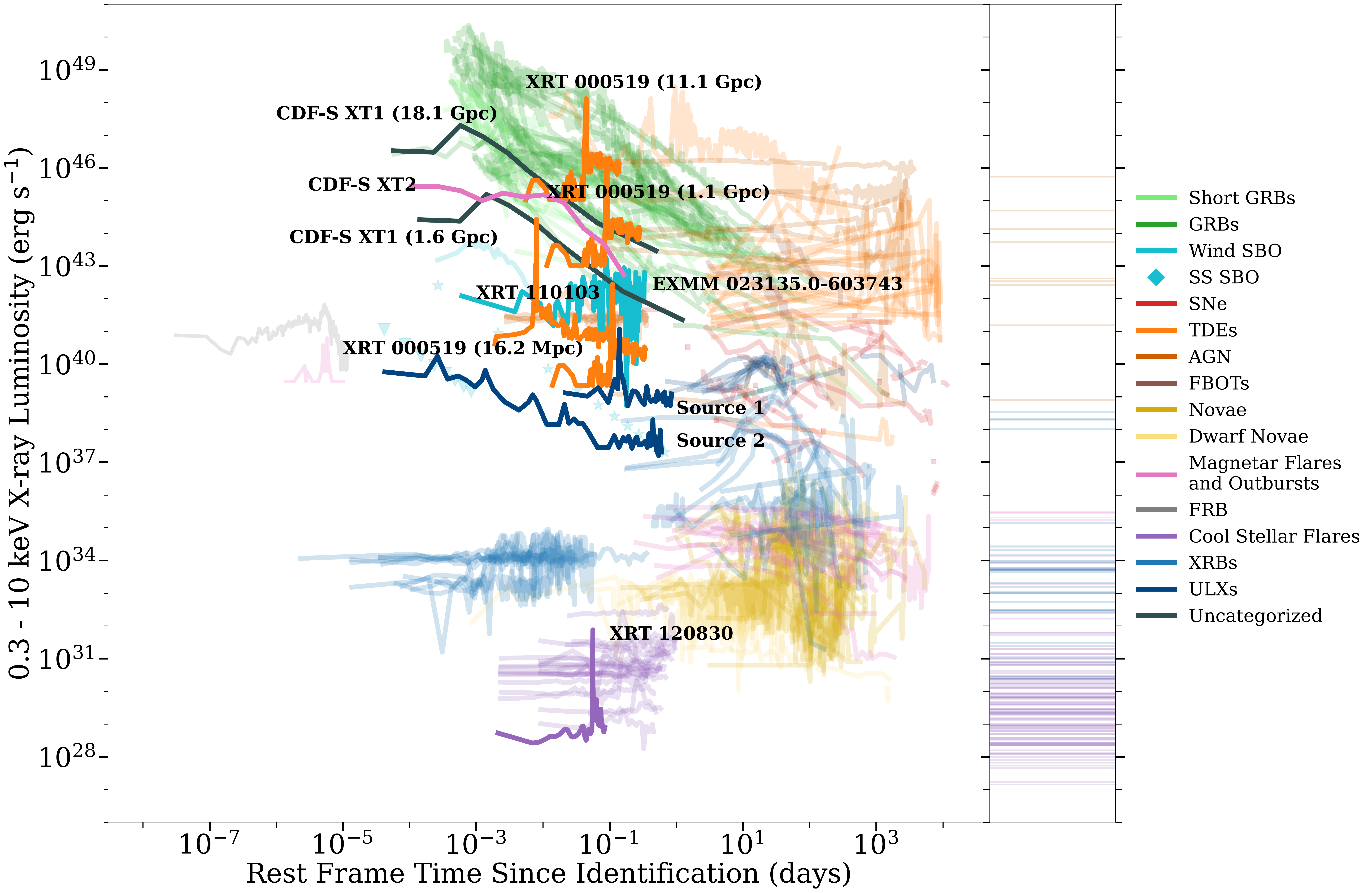}
\caption{We compare existing observations of transients with unclear/debated classification (Table \ref{tab:oddballs}) with our established X-ray phase space as described in Section \ref{subsec:unknown}. These signals are colored according to their preferred classification, though in cases where no one model is considered a better match (CDF-S XT1), we give them their own ``oddball'' color to differentiate them from the already classified transients in the phase space. For sources with uncertain distance estimates, each estimate is shown, with the relevant distance stated in parentheses. Included events are listed in Table \ref{tab:oddballs}. \label{fig:OddballDLPS}}
\end{figure*}

With the rise of time-domain astronomy, there has been a commensurate increase in opportunities to capture new types of transient/variable events that defy all known classification schemes. In some cases, these events have been discovered in archival data searches, thus preventing real time follow-up of these events outside the X-rays. This has practically prevented the identification of the true underlying nature of these new classes of events. In Figure \ref{fig:OddballDLPS}, we plot a selection of these yet-unidentified ``oddballs'' (\citealt{Jonker_2013}; \citealt{Glennie_2015}; \citealt{2016Natur.538..356I}; \citealt{Bauer_2017}; \citealt{2019Natur.568..198X}; \citealt{2020ApJ...898...37N}) to illustrate how they fit into the larger X-ray phase space. We include only those with known or estimated distances (assuming for the purposes of this case study that the X-ray phase space of transients for $z \leq 1$ is similar to the phase space of transients at all redshifts), and spectra for those observed outside of the 0.3 - 10 keV energy band to facilitate a k-correction.

Where multiple distance estimates are given, we include light curves at each of those distances to better fill out the \textit{uncategorized} X-ray phase space and demonstrate the potentially varied interpretations of these signals at different redshifts. Though they are sometimes referred to as fast X-ray transients (FXTs or FXRTs, see e.g. \citealt{2022AA...663A.168Q, 2023arXiv230413795Q} for a population-level examination), the light curves point to these transients having a variety of progenitors. This inhomogeneous class of transient events evolves on timescales of $\sim 10$s of seconds to days and spans roughly 21 dex in luminosity. 

Where the discovery papers have broadly speculated about the origin of these transients, we have colored the ``oddball'' light curves accordingly, allowing their position in the DLPS to discriminate between equally likely physical scenarios. In fact, this is an extension of the analysis done in \citet{Bauer_2017}, where they illustrate potential classifications by comparing CDF-S XT1 to light curves from already classified events. Where no one potential class is favored in the discovery paper we choose to leave the light curve \textit{uncategorized} in the X-ray phase space (CDF-S XT1), whereas we color those with a single (or preferred) proposed origin according to that theory (XRT 000519, XRT 120830, and XRT 110103, EXMM 023135.0-603743). \citet{Jonker_2013} prefers the (beamed) tidal disruption of a white dwarf by an intermediate mass black hole for XRT 000519, though our results suggest that an X-ray flash (as would be associated with a GRB) would also be reasonable. Taking XRT 000519 as potentially related to XRT 110103, \citet{Glennie_2015} suggest the same potential progenitors for that event. They also indicate that XRT 120830 seems consistent with a dwarf star flare, which is borne out by its position in the phase space. \citet{2022MNRAS.514..302E} examines the potential host galaxies of XRT 000519 and XRT 110103 in order to place constraints on the nature of the FXTs; though no potential host was detected for XRT 110103, XRT 000519 appears associated in projection with a distant galaxy candidate, seemingly favoring a beamed TDE or a binary neutron star merger like the one responsible for GW170817. 

Sources 1 and 2 \citep{2016Natur.538..356I} are ultraluminous X-ray outbursts in NGC 4636 and NGC 5128 respectively. Though their behavior is largely consistent with soft gamma repeaters or anomalous X-ray pulsars (their position in the DLPS matches the anticipated position of intermediate luminosity/duration magnetar flares and outbursts), the stellar populations of their hosts make this scenario unlikely. It is also possible that they are outbursts due to accretion onto neutron stars (though they are super-Eddington in this scenario, these events are somewhat shorter in duration than the other ULXs in the DLPS) or intermediate mass black holes. In the context of the duration-luminosity phase space, Sources 1 and 2 are also consistent with the anticipated position of smaller, lower energy stellar surface shock breakouts. CDF-S XT2 \citep{2019Natur.568..198X} has been identified as having emission consistent with a magnetar-driven outburst, potentially from a binary neutron star merger. Its position in the DLPS is strikingly similar to the included population of short gamma-ray bursts. \citet{2019Natur.568..198X} rule out long gamma-ray bursts and shock breakout-like events due to the luminosity and luminosity evolution and point out that a beamed TDE is also unlikely due to the short timescales on which that evolution occurs. Similarly, \citet{2020ApJ...898...37N} posit that EXMM 023135.0-603743 could be a shock breakout from a core-collapse supernova, a possibility which is supported by the light curve's position in the phase space, while also noting that it could be an AGN (within the DLPS, EXMM 023135 also overlaps almost entirely with the QPE GSN 069), a TDE, or even a late-time observation of a giant flare from a magnetar, though each of those scenarios is disfavored given other concurrent data \citep[][]{2020ApJ...898...37N}. 

Ultimately it is clear that, while the DLPS is not able to provide classification for transients/variables without input from the signal's spectral evolution and from other investigations that hint at the underlying mechanism, it is extremely useful to contextualize potential and preliminary classifications. As we see looking at the 16.2 Mpc XRT 000519 light curve and XRT 110103, their position in the phase space is apparently more consistent with an AGN/QPE than with a TDE, the potential confusion in classification stemming from the innate difficulty in distinguishing TDEs and AGN. For greater distances, the light curve characteristics of XRT 000519 seem to potentially align with a GRB-related X-ray flare. Similarly, while CDF-S XT1 has a myriad of potential progenitors (among them, an off-axis short GRB or a subluminous GRB, another white dwarf-intermediate mass black hole TDE; \citealt{Bauer_2017}), the light curve (assuming a distance of
18.1 Gpc) is nicely consistent with the subluminous GRBs in our sample.

\subsection{Discovery Space} \label{subsec:discovery}
\begin{figure*}[ht!]
\epsscale{1.2}
\plotone{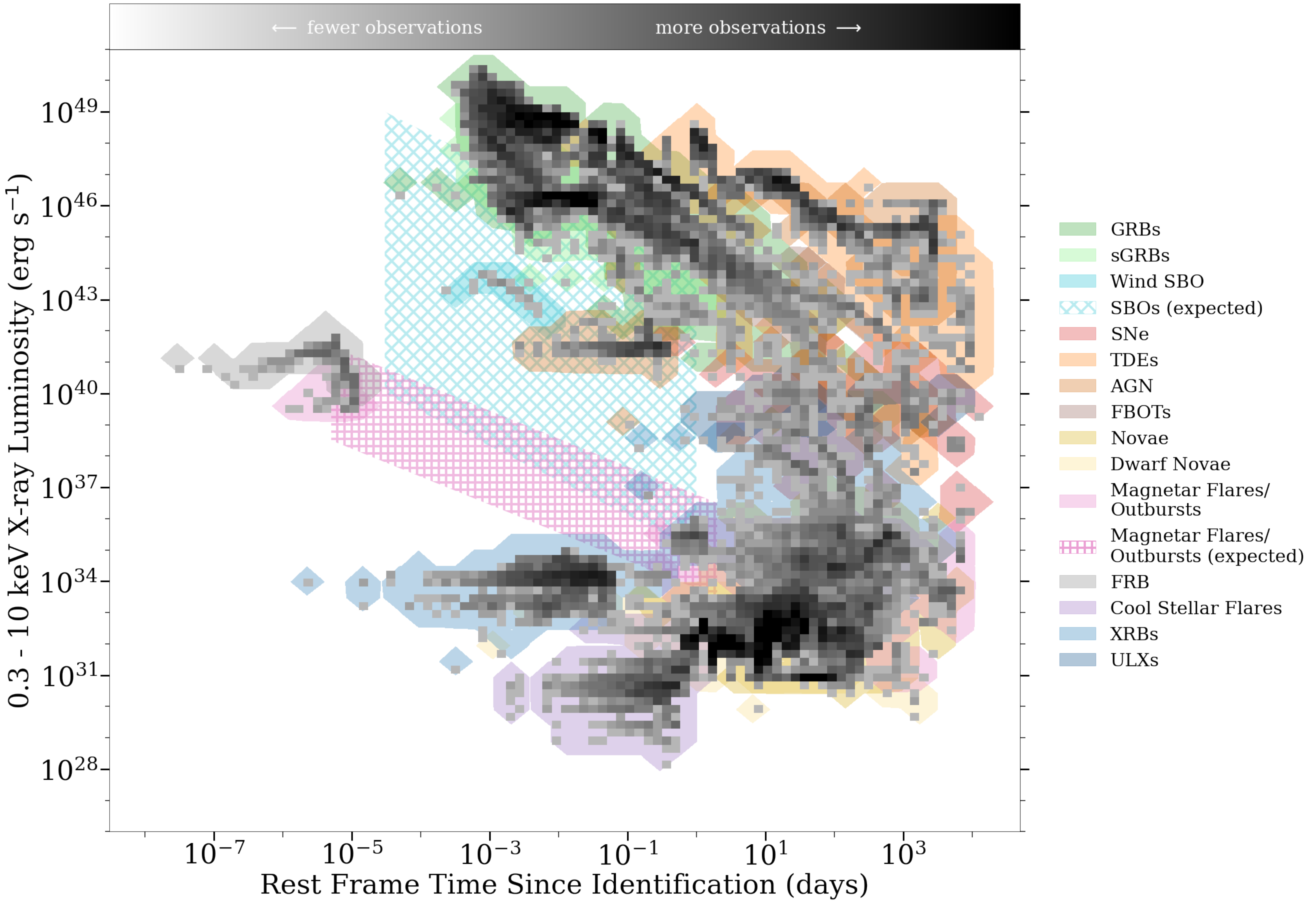}
\caption{The density of light curves in our phase space with the corresponding classes of transient underplotted; the colorbar is logarithmic and larger bins were used for the transient classes than for the overall observations. Though we only show the density of the representative data included in this paper (and so not the comprehensive density of \textit{all} observations in this energy band, though this sample should span a representative range in observed X-ray luminosity and duration), certain trends are notable that are generally relevant, including that the best sampled classes of transient are either Galactic phenomena (such as cool stellar flares or novae) or high luminosity extragalactic transients such as GRBs and short GRBs and that there is a paucity of observations of relatively short duration events at intermediate luminosities. We use hatches to mark the general region of the phase space where we would anticipate, but do not yet have, observations of magnetar flares and outbursts (pink) and shock breakouts from stellar explosions (light blue) among other events. We note that, though it is possible that the soft X-ray emission from giant flares of magnetars may be comparable in luminosity to what has been observed in the hard X-rays (peak $L_x \sim 10^{47}$ erg s$^{-1}$; \citealt{2005Natur.434.1098H}), given the paucity of soft X-ray observations of giant flares, we define the ``expected'' region of the DLPS for magnetar flares and outbursts based on available data.\label{fig:detdens}}
\end{figure*}

As we enter a new era in the search for/detection of X-ray transients and variables, due to both large time-domain surveys and next generation X-ray observatories, it is crucial to understand the observational restrictions that have inherently shaped our understanding of the high energy transient sky to now. In examining the phase space of existing detections, we find that while both the most luminous (largely extragalactic) and least luminous (largely Galactic) part of the phase space is well-populated at $t > 0.1$ days, intermediate luminosity phenomena (L$_x = 10^{34} - 10^{42}$ erg s$^{-1}$) represent a gap in the phase space. We thus identify L$_x = 10^{34} - 10^{42}$ erg s$^{-1}$ and $t = 10^{-4} - 0.1$ days as a key discovery phase space in transient X-ray astronomy (see Figure \ref{fig:detdens}). 

\begin{figure}[ht!]
\epsscale{1.2}
\plotone{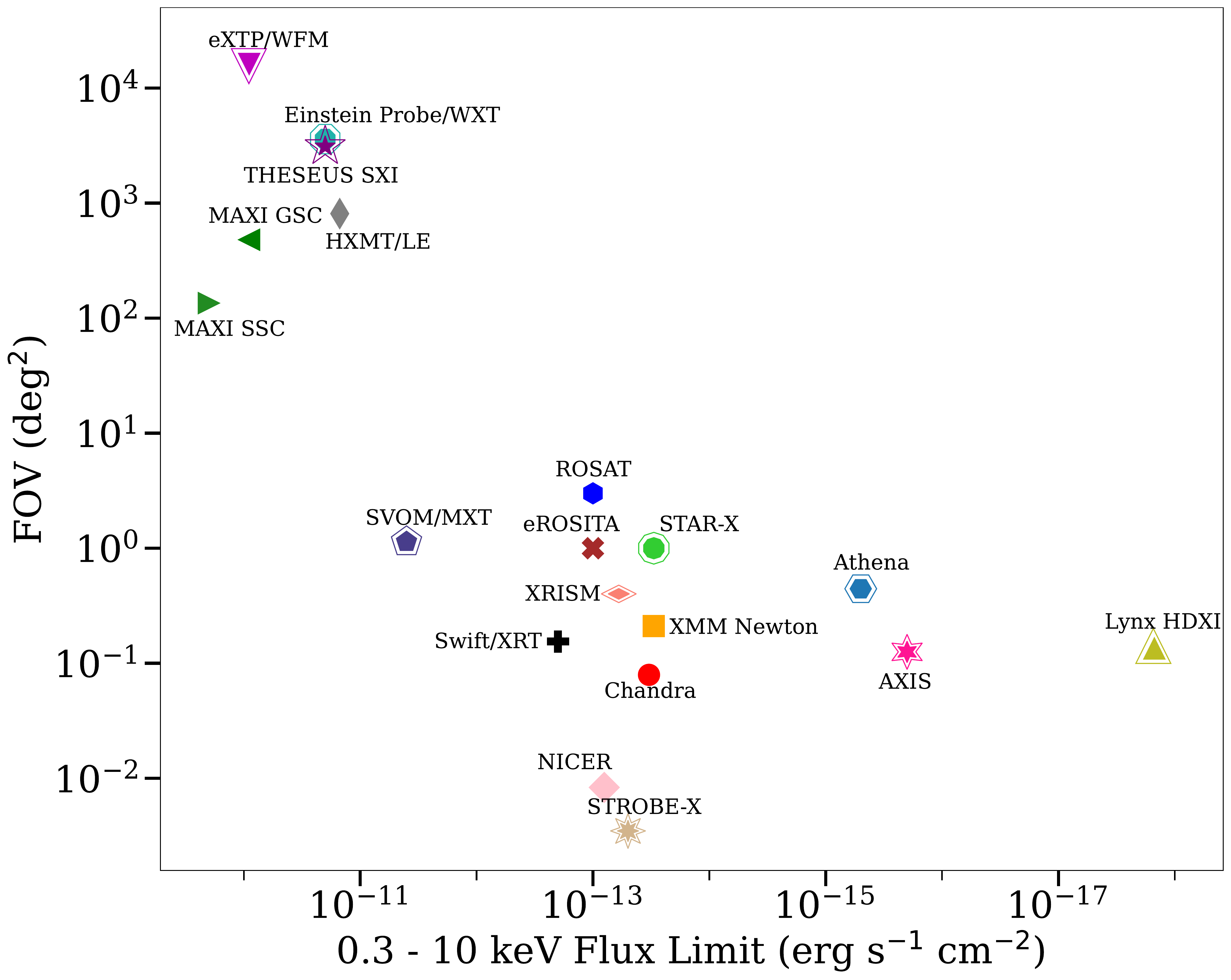}
\caption{Here we show the roughly inverse relation between instrument FOV and depth with a summary of these specifications for existing and planned/proposed X-ray missions (see Table \ref{tab:inst} for more details). We report the 1 ks 0.3-10 keV sensitivity. Upcoming/proposed instruments are highlighted by an additional marker outline. \label{fig:etendue}}
\end{figure}

The most obvious constraints are the sensitivity limits of current instruments and the difficulty of rapid response to a fleeting and intrinsically rare signal, which leave gaps in our phase space at low luminosities and short durations respectively. Due to inherent design constraints (see Figure \ref{fig:etendue}, discussed more in Section \ref{subsec:missions}) current instruments generally fall into one of two categories -- instruments that are likely to contribute to the serendipitous discovery of soft X-ray transients, which have limited sensitivity and instruments that allow for follow-up of event evolution down to very deep limits, which are extremely limited in their field of view.

Instruments with a wide field of view will serendipitously detect many more events than targeted instruments, contributing to the discovery of transient signals alongside survey instruments. Realistically, extremely short-duration events (on the order of seconds) will not be observed with any regularity without a new generation of wide field instruments.  This regime of extremely rapid events is already known to include FRB X-ray counterparts and their likely relatives, magnetar flares.

Target of Opportunity (ToO) protocols and other similar observational triggers play a role in successful follow-up of transitory signals. Greater efficiency in the form of fast re-pointings will also help push toward observation of extremely short-duration events; for instance, the robust \emph{Swift} ToO process is well-established. Automated follow-up is not restricted to the X-ray, with high-energy transient detections triggering radio observations \citep[e.g.][]{2013MNRAS.428.3114S, 2019PASA...36...46H}, as well.

Projects such as Exploring the X-ray Transient and variable Sky (EXTraS; \citealt{2021arXiv210502895D}) aim to address the gap in the short duration phase space at the algorithmic level, extracting previously unidentified signals and variability from existing XMM-Newton data (e.g., \citealt{2020ApJ...898...37N}). Efforts to rapidly disseminate information about detections like the Living \emph{Swift}-XRT Point Source catalog \citep[][]{2022arXiv220814478E} offer yet other opportunities for expedient analysis and follow-up. Similarly, it is possible that mining \emph{unrelated} X-ray observations (for example, those intended to study the hot halos of galaxies) for transients in real time provides another avenue for serendipitous detection.

More sensitive instruments are key for targeted follow-up. The next generation of highly sensitive soft X-ray missions will enable us to track the evolution of light curves to much later times/lower luminosities as they decay and will provide a broader understanding of transient populations, as in many cases, we are currently only meaningfully sampling the most luminous end of the population.

Figure \ref{fig:detdens} also reveals an under-sampled area of the phase space that we should aim to explore. On the interval $L_x = 10^{34} - 10^{42}$ erg s$^{-1}$ and with timescales between $10^{-4}$ and $\sim 0.1$ days, there is a clear gap in the phase space. This gap also corresponds to some known physical phenomena -- stellar surface shock breakouts (see Section \ref{subsec:SBO}) and the continuum behavior between magnetar flares and outbursts (see Section \ref{subsec:magnetar}). Efforts to expand observations in this regime should be motivated by the probable detection of these missing signals.

\begin{deluxetable*}{lcccc}
\tablenum{2}
\tablecaption{Soft X-ray imaging instrument performance parameters. \label{tab:inst}}
\tablewidth{\linewidth}
\tablehead{Instrument & Energy band (keV) & Flux limit\tablenotemark{a} (erg s$^{-1}$ cm$^{-2}$) & FOV & References}
\startdata
ROSAT/PSPC-C & 0.1 - 2.5 & $\sim 10^{-13}$ & 3\,deg$^2$ & \citealt{trumper_1990, 1996rouh.book.....B} \\
 &  &  &  &\citealt[][]{1999AAS..138..441G}\\
\hline
Chandra ACIS-S\tablenotemark{b} & 0.5 - 7 & $\sim 3 \times 10^{-14}$ & 16\farcm9$\times$ 16\farcm9 & \citealt{Chandra1}\\
Swift/XRT & 0.3 - 10 & $\sim 2 \times 10^{-13}$ & 23\farcm6 $ \times$ 23\farcm6 & \citealt[][]{2005SSRv..120..165B}\\
 &  &  &  & \citealt{2020ApJS..247...54E}\\
MAXI GSC & 2 - 30 & $\sim 9 \times 10^{-11}$ & 160\arcdeg$\times$ 3\arcdeg & \citealt{2010fym..confE..14S}\\ 
MAXI SSC & 0.5 - 12 & $\sim 2 \times 10^{-10}$ &  90\arcdeg$\times$ 1\fdg5 & \citealt{10.1093/pasj/62.6.1371}\\
XMM-Newton/EPIC-pn & 0.2 - 10 & $\sim 3\times10^{-14}$ & 27\farcm5 $ \times$ 27\farcm5 & \citealt{2001AA...365L..51W}\\
SRG/eROSITA & 0.2 - 8 & $\sim10^{-13}$ & 0.8\,deg$^2$ &\citealt{2012arXiv1209.3114M}\\
NICER & 0.2 - 12 & $\sim 8 \times 10^{-14}$ & 30\,arcmin$^2$ & \citealt[][]{2014SPIE.9144E..20A}\\
Insight-HXMT/LE & 0.9 - 12 & $\sim 1.5 \times 10^{-11}$ & 21$\times$(1\fdg6$\times$6\arcdeg), 7$\times$(4\arcdeg $\times$6\arcdeg), & \citealt{2020JHEAp..27...64L}\\
& & &2$\times$(50$\sim60$\arcdeg$\times$2$\sim$6\arcdeg)& \\
& & & or $\sim 810$ deg$^2$ total& \\
\hline
SVOM/MXT & 0.2 - 10 & $\sim 4 \times 10^{-12}$ & 64\arcmin $\times$ 64\arcmin & \citealt{2016arXiv161006892W} \\ 
XRISM/Xtend\tablenotemark{c} &0.4 - 13 & $\sim 6\times 10^{-14}$ & 38 \arcmin$\times$ 38 \arcmin & XRISM Team, \\
 & &  & & Private Communication\\
Athena/WFI & 0.2 - 15 & $\sim 5 \times 10^{-16}$ & 40\arcmin $\times$40\arcmin &\citealt{2012arXiv1207.2745B}\\
\hline
eXTP/WFM & 2 - 50 & $\sim 9 \times 10^{-11}$ & $\sim$ 180\arcdeg $\times$ 90\arcdeg & \citealt{2019SCPMA..6229502Z}\\
AXIS & 0.1 - 10 & $\sim 2 \times 10^{-16}$ & 144$\pi$ arcmin$^2$ & \citealt{2018SPIE10699E..29M}\\
Einstein Probe/WXT & 0.5 - 4 & $\sim 2 \times 10^{-11}$ & 3600 deg$^2$ & \citealt{2017symm.conf..247Y}\\
STAR-X & 0.5 - 6 & $\sim 3\times10^{-14}$ & 1 deg$^2$ & STAR-X Team,\\
&  &  & & Private Communication\\
STROBE-X & 0.2 - 12 & $\sim 5\times10^{-14}$ & 4$\pi$ arcmin$^2$ & \citealt{2019arXiv190303035R} \\
 &  &  &  &\citealt{2018CoSka..48..498M}\\
Lynx/HDXI & 0.2 - 10 & $\sim1.5 \times 10^{-18}$ & 22\arcmin $\times$ 22\arcmin & \citealt{Lynx_report}\\
THESEUS/SXI & 0.3 - 5 & $\sim 2 \times 10^{-11}$ & $\sim 0.5$\,sr & \citealt{2021ExA....52..183A}
\enddata
\tablecomments{We use the horizontal bars to differentiate between four categories of instrument, from top to bottom we list past instruments, currently operational instruments, instruments on future missions selected for launch, and  instruments on proposed missions. The list of proposed missions is not complete and it is provided to illustrate the range of capabilities of future experiments.}
\tablenotetext{a}{0.3-10 keV; all flux limits are k-corrected to our band-of-interest assuming a fiducial $\Gamma = 2$ spectrum. Flux limit is based on a 1\,ks exposure for instruments that do pointed observations. We note that for instruments designed for higher energy observations -- such as MAXI GSC or eXTP/WFM -- our estimated flux limit in the 0.3 - 10 keV energy band is less secure.}
\tablenotetext{b}{The reported Chandra 0.3 - 10 keV flux limit is estimated from recent observations}
\tablenotetext{c}{We take the full-band Suzaku/XIS flux limit from \citet{2008PASJ...60S..49M}, given its sensitivity is roughly comparable to that of XRISM/Xtend (XRISM Team, Private Communication).}
\end{deluxetable*}

\subsection{Rates of transient discovery}\label{subsec:missions}
There is a well-known trade-off between instrument FOV and sensitivity, as shown in Figure \ref{fig:etendue}, using specifications from currently operating and proposed missions: larger FOVs tend to correlate with lower sensitivities.

Making the reasonable assumption that we have already observed, and included in our DLPS, the most luminous events from each subclass of transient/variable, we can decouple the advantages of increased FOV and increased sensitivity with transient peak luminosities and volumetric rates from \citet{2016MNRAS.455..859S}, \citet{2020ApJ...895L..23C}, \citet{2021ARAA..59..155M}, and \citet{2022ApJ...932...10G}, separately examining the importance of wide-field instruments and extremely sensitive instruments.   For the GRBs (both long and short), we apply a beaming correction factor based on a conservative jet opening angle of $\sim 3^{\circ}$ \citep[][]{2015ApJ...815..102F, 2022arXiv221005695R}. Calculating the maximum distance out to which each class of transient can be observed based on instrument sensitivity, $d_\mathrm{L} = \sqrt{L_\mathrm{max}/(4\,\pi\times\mathrm{sensitivity})}$, we can use a three-dimensional observing volume, defined as $\frac{\mathrm{FOV}}{\Omega} \times \frac{4}{3}\pi \, \mathrm{d_{com}}^3$ (where FOV is the instrument field of view, $\Omega$ is the solid angle of the sky, and $\mathrm{d_{com}}$ is the maximum comoving distance out to which each class of transients can be observed inferred from $d_\mathrm{L}$ and the cosmology in Section \ref{sec:intro}), to ascertain the number of anticipated transient observations per year for instruments of varying sensitivity and FOV.

\begin{figure*}[ht!]
\epsscale{1.2}
\plotone{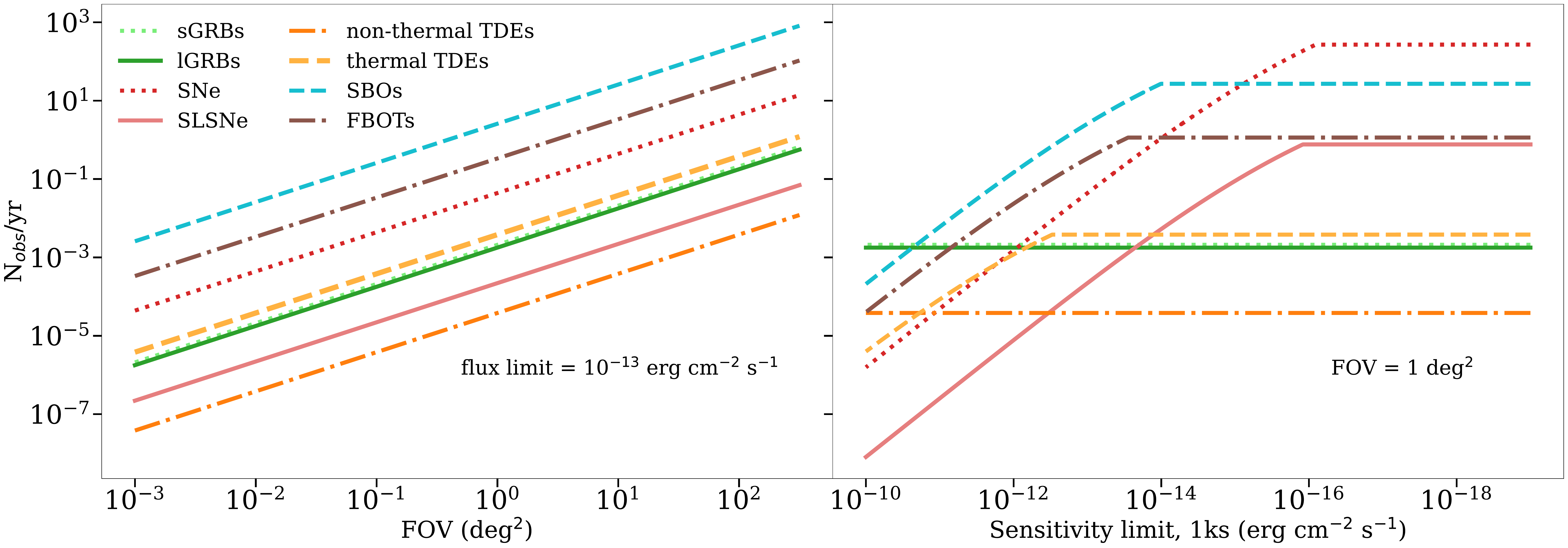}
\caption{To decouple the advantages of increased field of view and sensitivity, we show extragalactic transient rates \citep[][]{2016MNRAS.455..859S, 2020ApJ...895L..23C, 2021ARAA..59..155M, 2022ApJ...932...10G} as a function of FOV with a fixed flux limit (left, sensitivity $ = 10^{-13}$ erg s$^{-1}$ cm$^{-2}$) and the same rates as a function of sensitivity with a fixed field of view (right, FOV $ = 1$ deg$^2$). As in the rest of this work, we limit the rates to $z \le 1$ -- which is why the number of observations per year eventually flattens with increasing sensitivity. For luminous sources, like those included here, substantial increases in the number of events detected within $z \lesssim 1$ will primarily come from instruments with increased field of view.
\label{fig:sensFOV}}
\end{figure*}

We can isolate the impact of increased FOV by effectively marginalizing over sensitivity and looking at the number of observations as a function of field of view. We choose a representative sensitivity of $10^{-13}$ erg s$^{-1}$ cm$^{-2}$, corresponding to the median flux limit of the instruments listed in Table \ref{tab:inst}, and assume a static pointing, so that the importance of signals serendipitously falling in that FOV is clear. We show this plot on the left side of Figure \ref{fig:sensFOV}. Similarly, we can look at the importance of increased sensitivity by selecting a representative FOV (1 deg$^2$, corresponding to the median FOV among instruments in Table \ref{tab:inst}) and plotting the number of detections per class of transient as a function of sensitivity. This is shown on the right side of Figure \ref{fig:sensFOV}. As in the rest of the paper, we limit our events and observing depths to the redshift range in which their rates are well-constrained ($z \leq 1$). It is then apparent that, for detection of bright sources like GRBs, TDEs, SBOs, and FBOTs within $z = 1$, there is little, if anything, to be gained by improving instrument sensitivity beyond a 1 ks flux limit $\sim 10^{-14}$ erg s$^{-1}$ cm$^{-2}$; instead, wide FOV instruments will be critical to the discovery and observation of these events.

We can also use each class's maximum luminosity along with the instrument sensitivity to determine the distance out to which each transient/variable event can be observed. We take the maximum luminosity of the transient in the GRB, SBO, SN, TDE, CV, magnetar flare/outburst, cool stellar flare, XRB, ULX, FBOT, and FRB categories to represent the most luminous end of their respective distributions. We then apply the 0.3-10 keV flux limit (as in Figure \ref{fig:etendue} and Table \ref{tab:inst}) to determine the luminosity distance out to which the transient can be detected, from which we estimate the comoving distance.

\begin{figure*}[ht!]
\epsscale{1.}
\plotone{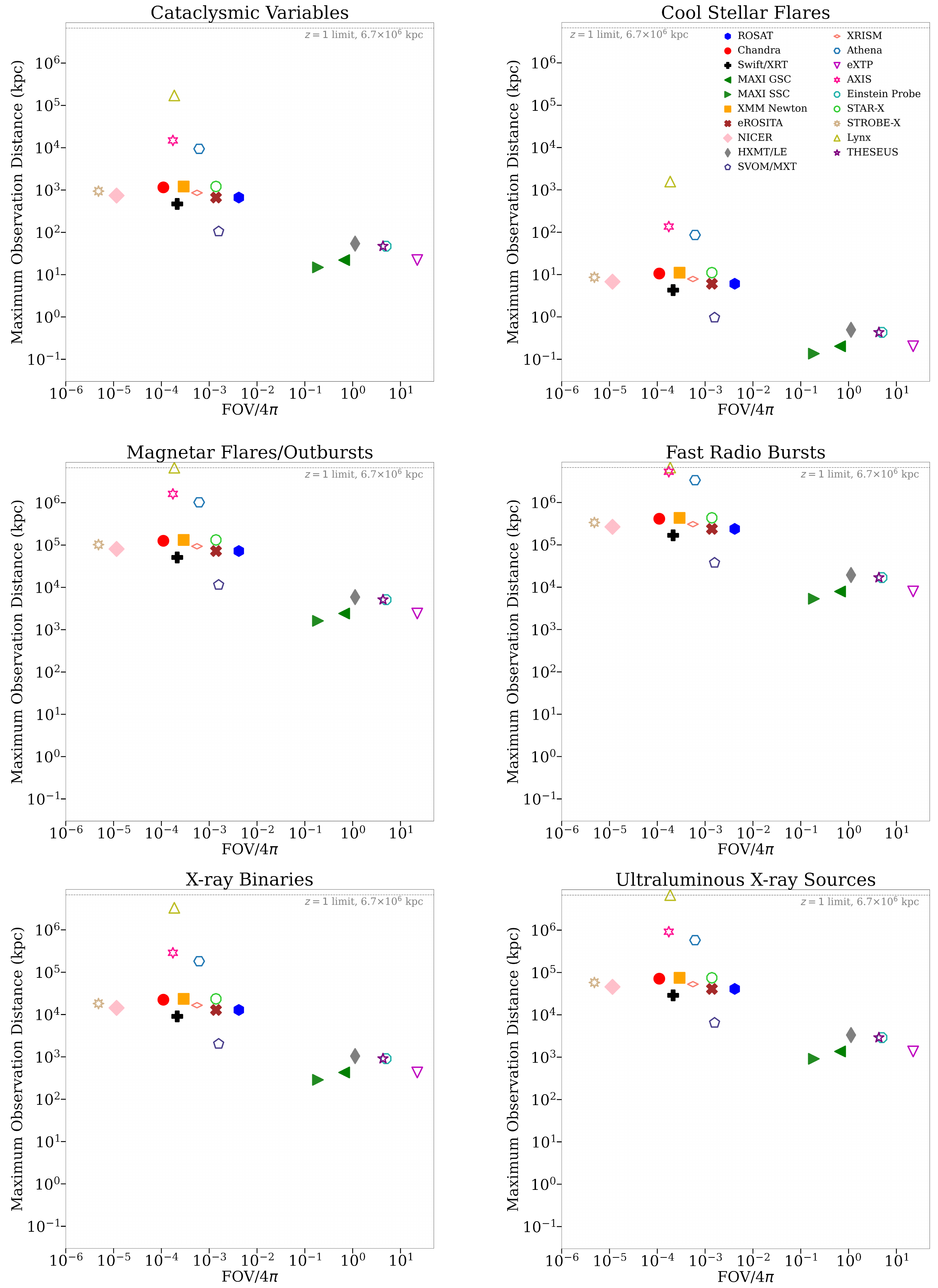}
\caption{The maximum distance out to which each instrument can observe different classes of transient/variable vs. the normalized FOV (by the area of the sky). As in the rest of this work, we limit to $z \leq 1$ ($d_{L} = 6.7\times10^6$ kpc), which is shown by the gray dashed line. For less luminous classes of transient, the improved sensitivity from proposed/planned instruments (shown with open markers) represents an increase of up to $\sim$ 3 orders of magnitude in the distance out to which these events can be detected and observed. This significantly enlarges the potential observing volume for these signals.\label{fig:galacticObsVolume}}
\end{figure*}

For Galactic transients (i.e., CVs, magnetar flares/outbursts, and cool stellar flares) and Galactic/extragalactic transients (i.e., XRBs, ULXs, and FRBs\footnote{Though the only FRB with an apparent X-ray counterpart (FRB200428) is a Galactic source, the population of observed fast radio bursts is largely extragalactic.}), we show the luminosity distance out to which various transients can be observed vs. the fraction of the sky covered instantaneously by the instrument FOV in Figure \ref{fig:galacticObsVolume}. As in the rest of this work, we limit instrument depth to a maximum luminosity distance that corresponds to $z = 1$. Needless to say, the qualitative trends captured by our Galactic and Galactic/extragalactic transient observing volume plot (Figure \ref{fig:galacticObsVolume}) translate to the behavior of exclusively extragalactic sources, with targeted instruments being more conducive to observing distant phenomena, while wide-field instruments have considerably shallower sky coverage.

\begin{figure*}[ht!]
\epsscale{1.2}
\plotone{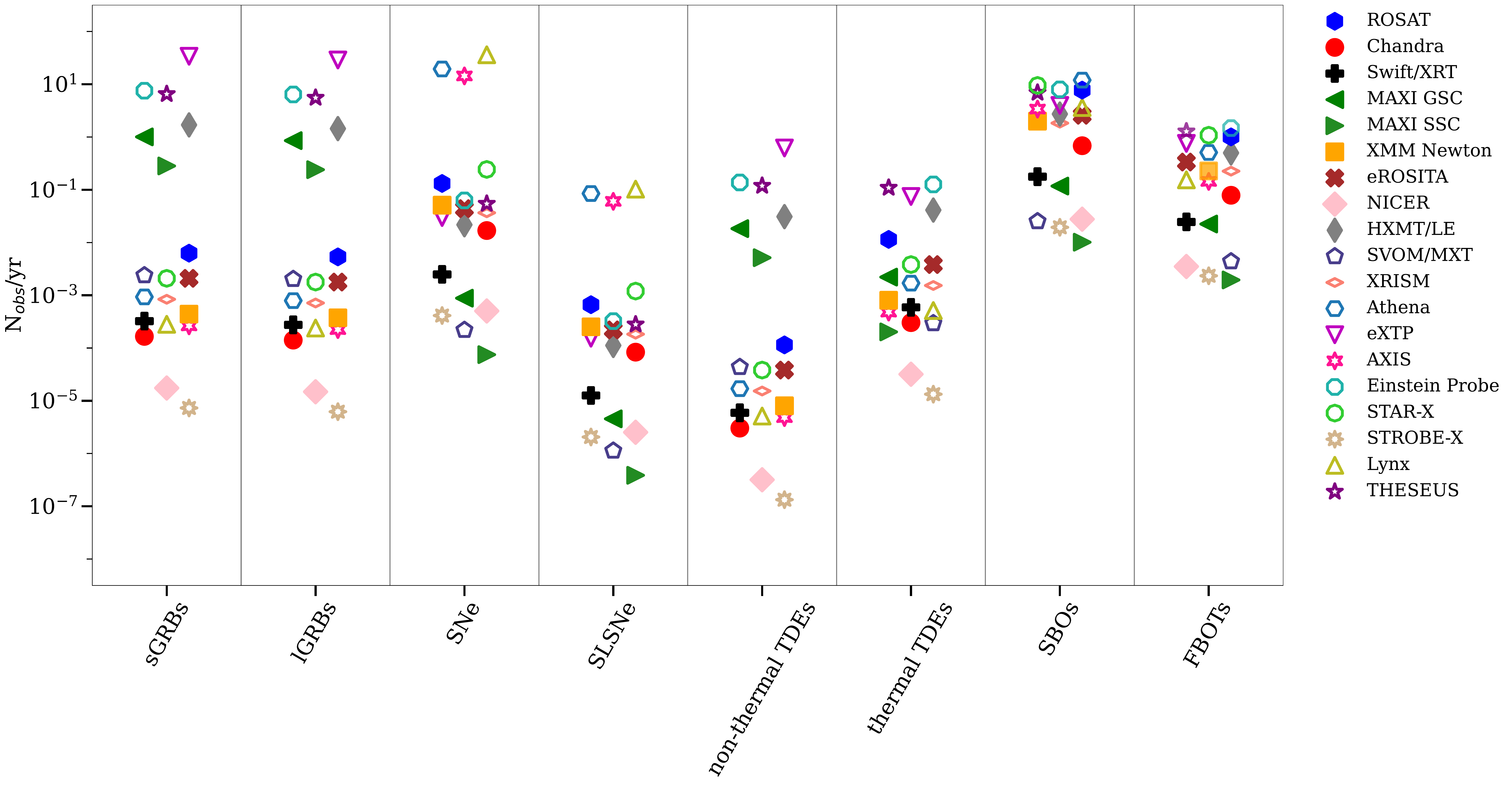}
\caption{Estimates of the number of serendipitous observations expected per year for transients with a variety of different instruments based on the volumetric rates of the phenomena and the observing volume of each instrument. The comoving depth out to which each instrument can observe each class of phenomenon is calculated based on the peak observed luminosity of each phenomenon, making the assumption that we have already detected the most intrinsically luminous signal from each type of transient. We take rates from \citet{2016MNRAS.455..859S},  \citet{2020ApJ...895L..23C}, \citet{2021ARAA..59..155M}, and \citet{2022ApJ...932...10G}. For the GRBs, both long and short, we apply a beaming correction assuming a jet opening angle $\sim 3^{\circ}$. See Table \ref{tab:rates} for more details.
As with the rest of the paper, the number of expected observations per year is quoted out to $z = 1$, corresponding to $d_L \sim$ 6700 Mpc or $d_{\rm{com}} \sim 3350$ Mpc. For each class of transient shown here, next-generation instruments (open markers) represent an increase in the anticipated number of events observed per year. In the case of more intrinsically luminous transients, this is due to increased FOV, while less luminous classes benefit more from improved sensitivity (see Figure \ref{fig:sensFOV}).
\label{fig:Nobs}}
\end{figure*}

For extragalactic transients (i.e., GRBs, SBOs, SNe, FBOTs, TDEs, and AGN), we instead examine an estimate of the number of observed events per year for each class of phenomena broken down by instrument. Using the three-dimensional observing volumes for each instrument included in Table \ref{tab:inst} and the same volumetric rates as in Figure \ref{fig:sensFOV}, we construct Figure \ref{fig:Nobs}. The rates shown in Figure \ref{fig:Nobs} are upper limits based on the most luminous event in each class. In Table \ref{tab:rates}, we report these upper limits in addition to lower limits (based on the least luminous observation in each extragalactic class in our DLPS). Both Figure \ref{fig:Nobs} and Table \ref{tab:rates} report anticipated serendipitous detection rates, assuming the instrument maintains a single static pointing on the sky and that each transient is observed as it undergoes a flare. This allows us to more readily compare instrument specifications without accounting for observing strategy. In reality, the observed rates may be much higher, particularly for survey instruments and classes of transient that can be observed months after the initial outburst \citep[e.g.,][]{2021MNRAS.508.3820S}.

\begin{deluxetable*}{lcccccccc}
\tablenum{3}
\tablecaption{Estimated rate (in yr$^{-1}$) of serendipitous detections of extragalactic transients in a single pointing of each soft X-ray instrument. \label{tab:rates}}
\tablewidth{\linewidth}
\tablehead{Instrument & sGRBs & lGRBs & SNe & SLSNe & non-thermal TDEs & thermal TDEs & SBOs & FBOTs}
\renewcommand{\arraystretch}{1.4}
\startdata
ROSAT & {$6.28\times10^{-3} \atop 8.98\times10^{-11}$} & {$5.34 \times 10^{-3} \atop 2.19 \times 10^{-6}$} & {$0.13 \atop 6.32 \times 10^{-13}$} & {$6.63 \times 10^{-4} \atop 9.96 \times 10^{-7}$} & {$1.15 \times 10^{-4} \atop 8.28 \times 10^{-8}$} & {$1.15\times10^{-2} \atop 1.59\times10^{-12}$} & {$7.74 \atop 2.85\times10^{-2}$} & {$1.00 \atop 3.62\times10^{-8}$} \\
\hline
Chandra & {$1.66\times10^{-4} \atop 1.25\times10^{-11}$} & {$1.41\times10^{-4} \atop 2.73 \times 10^{-7}$}& {$1.69 \times 10^{-2} \atop 8.81 \times 10^{-14}$} & {$8.35 \times 10^{-5} \atop 1.37 \times 10^{-7}$} & {$3.03 \times 10^{-6} \atop 1.01 \times 10^{-8}$} & {$3.03\times10^{-4} \atop 2.22\times10^{-13}$} & {$0.68 \atop 3.57\times10^{-3}$} & {$7.86 \times 10^{-2} \atop 5.04\times10^{-9}$} \\
Swift/XRT & {$3.24\times10^{-4} \atop 1.64\times10^{-12}$} & {$2.75\times10^{-4} \atop 4.18\times10^{-8}$}& {$2.49 \times 10^{-3} \atop 1.15 \times 10^{-14}$} & {$1.26 \times 10^{-5} \atop 1.83 \times 10^{-8}$} & {$5.91 \times 10^{-6} \atop 1.60 \times 10^{-9}$} & {$5.91\times10^{-4} \atop 2.91\times10^{-14}$} & {$0.18 \atop 5.43\times10^{-4}$} & {$2.46 \times 10^{-2} \atop 6.62\times10^{-10}$} \\
MAXI GSC & {$1.00 \atop 5.35 \times 10^{-13}$} & {$0.854 \atop 1.53\times10^{-8}$} & {$8.81 \times 10^{-4} \atop 3.68 \times 10^{-15}$} & {$4.55 \times 10^{-6} \atop 6.01 \times 10^{-9}$} & {$1.83 \times 10^{-2} \atop 5.98 \times 10^{-10}$} & {$2.21 \times 10^{-3} \atop 9.46 \times 10^{-15}$} & {$0.12 \atop 1.97\times10^{-4}$} & {$2.24 \times 10^{-2} \atop 2.16\times10^{-10}$} \\
MAXI SSC & {$0.283 \atop 4.54\times10^{-14}$} & {$0.240 \atop 1.39\times10^{-9}$} & {$7.49 \times 10^{-5} \atop 3.23 \times 10^{-16}$} & {$3.87 \times 10^{-7} \atop 5.10 \times 10^{-10}$} & {$5.16 \times 10^{-3} \atop 5.09 \times 10^{-11}$} & {$2.02 \times 10^{-4} \atop 8.03 \times 10^{-16}$} & {$1.01\times10^{-2} \atop 1.68\times10^{-5}$} & {$.95 \times 10^{-3} \atop 1.83\times10^{-11}$} \\
XMM Newton & {$4.40 \times 10^{-4} \atop 3.81 \times 10^{-11}$} & {$3.74 \times 10^{-4} \atop 8.22 \times 10^{-7}$} & {$5.10 \times 10^{-2} \atop 2.69 \times 10^{-13}$} & {$2.52 \times 10^{-4} \atop 4.19 \times 10^{-7}$} & {$8.02 \times 10^{-6} \atop 3.04 \times 10^{-8}$} & {$8.02\times10^{-4} \atop 6.79\times10^{-13}$} & {$2.00 \atop 1.08\times10^{-2}$} & {$0.23 \atop 1.54\times10^{-8}$} \\
eROSITA & {$2.09\times10^{-3} \atop 2.99\times10^{-11}$} & {$1.78\times10^{-3} \atop 7.29 \times 10^{-7}$} & {$4.39 \times 10^{-2} \atop 2.11 \times 10^{-13}$} & {$2.21 \times 10^{-4} \atop 3.32 \times 10^{-7}$} & {$3.82 \times 10^{-5} \atop 2.76 \times 10^{-8}$} & {$3.82\times10^{-3} \atop 5.31\times10^{-13}$} & {$2.58 \atop 9.49\times10^{-3}$} & {$0.334 \atop 1.21\times10^{-8}$} \\
NICER & {$1.74\times10^{-5} \atop 3.48\times10^{-13}$} & {$1.48\times10^{-5} \atop 8.33 \times 10^{-9}$} & {$5.04 \times 10^{-4} \atop 2.45 \times 10^{-15}$} & {$2.53 \times 10^{-6} \atop 3.86 \times 10^{-9}$} & {$3.18 \times 10^{-7} \atop 3.14 \times 10^{-10}$} & {$3.18\times10^{-5} \atop 6.19\times10^{-15}$} & {$2.77\times10^{-2} \atop 1.09\times10^{-4}$} & {$3.51 \times 10^{-3} \atop 1.41\times10^{-10}$} \\
Insight-HXMT/LE & {$1.70 \atop 1.33\times10^{-11}$} & {$1.44 \atop 3.76 \times 10^{-7}$} & {$2.17 \times 10^{-2} \atop 9.25 \times 10^{-14}$} & {$1.12 \times 10^{-4} \atop 1.49 \times 10^{-7}$} & {$3.09 \times 10^{-2} \atop 1.47 \times 10^{-8}$} & {$4.12 \times 10^{-2} \atop 2.35 \times 10^{-13}$} & {$2.73 \atop 4.85\times10^{-3}$} & {$0.50 \atop 5.35\times10^{-9}$} \\
\hline
SVOM/MXT & {$2.38\times10^{-3} \atop 1.35\times10^{-13}$} & {$2.03\times01^{-3} \atop 3.78 \times 10^{-9}$} &  {$2.19 \times 10^{-4} \atop 9.50 \times 10^{-16}$} & {$1.13 \times 10^{-6} \atop 1.52 \times 10^{-9}$} & {$4.35 \times 10^{-5} \atop 1.47 \times 10^{-10}$} & {$1.73\times10^{-4} \atop 2.39\times10^{-15}$} & {$2.52\times10^{-2} \atop 4.89\times10^{-5}$} & {$4.38 \times 10^{-3} \atop 5.46\times10^{-11}$} \\
XRISM/Xtend & {$8.40\times10^{-4} \atop 2.58\times10^{-11}$} & {$7.14\times10^{-4} \atop 6.01 \times 10^{-7}$} & {$3.66 \times 10^{-2} \atop 1.82 \times 10^{-13}$} & {$1.83 \times 10^{-4} \atop 2.85 \times 10^{-7}$} & {$1.53 \times 10^{-5} \atop 2.26 \times 10^{-8}$} & {$2.94 \times 10^{-4} \atop 2.39 \times 10^{-15}$} & {$1.83 \atop 7.84\times10^{-3}$} & {$0.22 \atop 1.04\times10^{-8}$} \\
Athena WFI & {$9.31\times10^{-4} \atop 3.50\times10^{-8}$} & {$7.91\times10^{-4} \atop 2.55 \times 10^{-4}$} & {$19.5 \atop 2.65 \times 10^{-10}$} & {$8.49 \times 10^{-2} \atop 3.38 \times 10^{-4}$} & {$1.70 \times 10^{-5} \atop 8.20 \times 10^{-6}$} & {$1.70\times10^{-3} \atop 6.58\times10^{-10}$} & {$12.0 \atop 3.46$} & {$0.51 \atop 1.42\times10^{-5}$} \\
\hline
eXTP/WFM & {$33.9 \atop 1.80\times10^{-11}$} & {$28.8 \atop 5.16 \times 10^{-7}$} & {$2.97 \times 10^{-2} \atop 1.24 \times 10^{-13}$} & {$1.53 \times 10^{-4} \atop 2.03 \times 10^{-7}$} & {$0.62 \atop 2.02 \times 10^{-8}$} & {$7.46 \times 10^{-2} \atop 3.19 \times 10^{-13}$} & {$3.97 \atop 6.66\times10^{-3}$} & {$0.76 \atop 7.28\times10^{-9}$} \\
AXIS & {$2.63\times10^{-4} \atop 3.75\times10^{-8}$} & {$2.24\times10^{-4} \atop 1.77 \times 10^{-4}$} & {$14.4\atop 2.96 \times 10^{-10}$} & {$6.03 \times 10^{-2} \atop 3.37 \times 10^{-4}$} & {$4.80\times10^{-6} \atop 4.80\times10^{-6}$} & {$4.80\times10^{-4} \atop 7.28\times10^{-10}$} & {$3.38 \atop 2.42$} & {$0.14 \atop 1.53\times10^{-5}$} \\
Einstein Probe/WXT & {$7.54 \atop 3.83\times10^{-11}$} & {$6.41 \atop 1.09 \times 10^{-6}$} & {$6.28 \times 10^{-2} \atop 2.68 \times 10^{-13}$} & {$3.24 \times 10^{-4} \atop 4.30 \times 10^{-7}$} & {$0.14 \atop 4.24 \times 10^{-8}$} & {$0.13 \atop 6.77 \times 10^{-13}$} & {$7.99 \atop 1.40\times10^{-2}$} & {$1.48 \atop 1.54\times10^{-8}$}\\
STAR-X & {$2.09\times10^{-3} \atop 1.81\times10^{-10}$} & {$1.78\times10^{-3} \atop 3.91 \times 10^{-6}$} & {$0.24 \atop 1.28 \times 10^{-12}$} & {$1.20 \times 10^{-3} \atop 1.99 \times 10^{-6}$} & {$3.82 \times 10^{-5} \atop 1.45 \times 10^{-7}$} & {$3.82\times10^{-3} \atop 3.23\times10^{-12}$} & {$9.50 \atop 5.13\times10^{-2}$} & {$1.08 \atop 7.32\times10^{-8}$} \\
STROBE-X & {$7.31\times10^{-6} \atop 2.95\times10^{-13}$} & {$6.21\times10^{-6} \atop 6.75 \times 10^{-9}$} & {$4.13 \times 10^{-4} \atop 2.08 \times 10^{-15}$} & {$2.06 \times 10^{-6} \atop 3.26 \times 10^{-9}$} & {$1.33 \times 10^{-7} \atop 2.52 \times 10^{-10}$} & {$1.33\times10^{-5} \atop 5.24\times10^{-15}$} & {$1.94\times10^{-2} \atop 8.81\times10^{-5}$} & {$2.34\times10^{-3} \atop 1.19\times10^{-10}$} \\
Lynx/HDXI & {$2.82\times10^{-4} \atop 2.62\times10^{-5}$} & {$2.39\times10^{-4} \atop 2.39\times10^{-4}$} & {$36.2 \atop 4.85 \times 10^{-7}$} & {$0.10 \atop 9.97 \times 10^{-2}$} & {$5.14\times10^{-6} \atop 5.14\times10^{-6}$} & {$5.14\times10^{-4} \atop 9.33\times10^{-7}$} & {$3.62 \atop 3.62$} & {$0.15 \atop 1.13\times10^{-2}$} \\
THESEUS/SXI & {$6.51 \atop 3.31\times10^{-11}$} & {$5.54 \atop 9.39 \times 10^{-7}$} & {$5.43 \times 10^{-2} \atop 2.32 \times 10^{-13}$} & {$2.80 \times 10^{-4} \atop 3.71 \times 10^{-7}$} & {$0.12 \atop 3.67 \times 10^{-8}$} & {$0.11 \atop 5.85\times10^{-13}$} & {$6.90 \atop 1.21\times10^{-2}$} & {$1.28 \atop 1.33\times10^{-8}$} \\
\enddata
\tablecomments{For each instrument we show an upper (top) and lower (bottom) limit for the serendipitous observation rate, assuming that the instrument remains pointed at the same portion of the sky and takes 1ks exposures}. The upper limit is based on the observing volume calculated for the most luminous observation of an event in that class, while the lower limit is based on the observing volume calculated for the least luminous observation of an event in that class. As in Table \ref{tab:inst}, we separate past, present, planned, and proposed instruments with horizontal lines.
\end{deluxetable*}

\section{Conclusion}
With the immense promise of next generation X-ray instruments on the horizon and community investment in large time-domain surveys, many more X-ray transients will be detected and studied in the coming years. We constructed a set of observational X-ray phase space plots from 284 light curves of 221 objects, which show distinctions between different transient and variable phenomena and highlight the luminosity evolution of these events. We included light curves of gamma-ray burst afterglows, supernovae, shock breakouts, tidal disruption events and active galactic nuclei, fast blue optical transients, cataclysmic variables, magnetar flares/outbursts and fast radio bursts, cool stellar flares, X-ray binary outbursts, and ultraluminous X-ray sources comprised of observations from a range of telescopes (see Table \ref{tab:class} for a full list). The X-ray duration-luminosity phase space can be used to help disambiguate the nature of newly observed signals by placing them in context (even before spectroscopic or multi-wavelength follow-up, as demonstrated in Section \ref{subsec:unknown}) and to point out sparse areas of the phase space that should be the focus of future exploration.

As expected, the phase space is less populated at extremely low luminosities and extremely short durations, given the limitations of current instruments and the trade-off between FOV and sensitivity in instrument design. More sensitive imagers will provide better insight into less luminous events, but wide-field imagers will be necessary to serendipitously capture the most ephemeral signals, like those of candidate FRB counterparts. There is another, less intuitive gap in the phase space around $L_x = 10^{34} - 10^{42}$ erg s$^{-1}$ and duration $10^{-4} - 0.1$ days. We expect this part of the X-ray phase space to include both SBOs and magnetar flares, both of which are classes of transient that have a relative paucity of observations. Additional observations targeting this part of the phase space will not only increase the studied population of known classes of transient, but will potentially uncover yet-unidentified signals, as well.

\acknowledgments
 The authors thank the anonymous referees for
comments and suggestions that significantly improved the
manuscript, as well as Laura Chomiuk, Irina Zhuravleva, and Magaretha Pretorius for helpful discussions. A.P. thanks Andrey Kravtsov for the suggestion to make a preliminary classifier available. K.L.P. acknowledges support from the UK Space Agency.
This work made use of data supplied by the UK Swift Science Data Centre at the University of Leicester.
R.M. acknowledges partial support by the National Science Foundation under Grant No. AST-2221789 and AST-2224255, by the Heising-Simons Foundation under grant \# 2021-3248. 
G.V. acknowledges support by Hellenic Foundation for Research and Innovation (H.F.R.I.) under the ``3rd Call for H.F.R.I. Research Projects to support Postdoctoral Researchers'' through the project ASTRAPE (Project ID 7802).

This research has made use of MAXI data provided by RIKEN, JAXA and the MAXI team. 

\software{AstroPy \citep{astropy:2013, astropy:2018}, Matplotlib \citep[][]{Hunter:2007}, NumPy \citep[][]{harris2020array}, pandas \citep[][]{reback2020pandas, mckinney-proc-scipy-2010}, PIMMS \citep{1993Legac...3...21M}, scikit-learn \citep[][]{scikit-learn}, SciPy \citep[][]{2020SciPy-NMeth}, WebPlotDigitizer \citep{WebPlotDigitizer}}

\clearpage
\appendix
\section{Data} \label{appA}
We list here all of the events included in the paper. For each event, we also provide coordinates, distance, and references, and, where applicable, we provide sub-classification. For GRBs, we also list redshift and $T_{90}$. These data are available on GitHub (see Section \ref{sec:data}), with a few limited exceptions, which are marked clearly in the tables below. Quoted distances are luminosity distances for the cosmology indicated in Section \ref{sec:intro}.

\section{The true observer's phase space: flux vs. duration}
\begin{figure*}[ht!]
\epsscale{1.}
\plotone{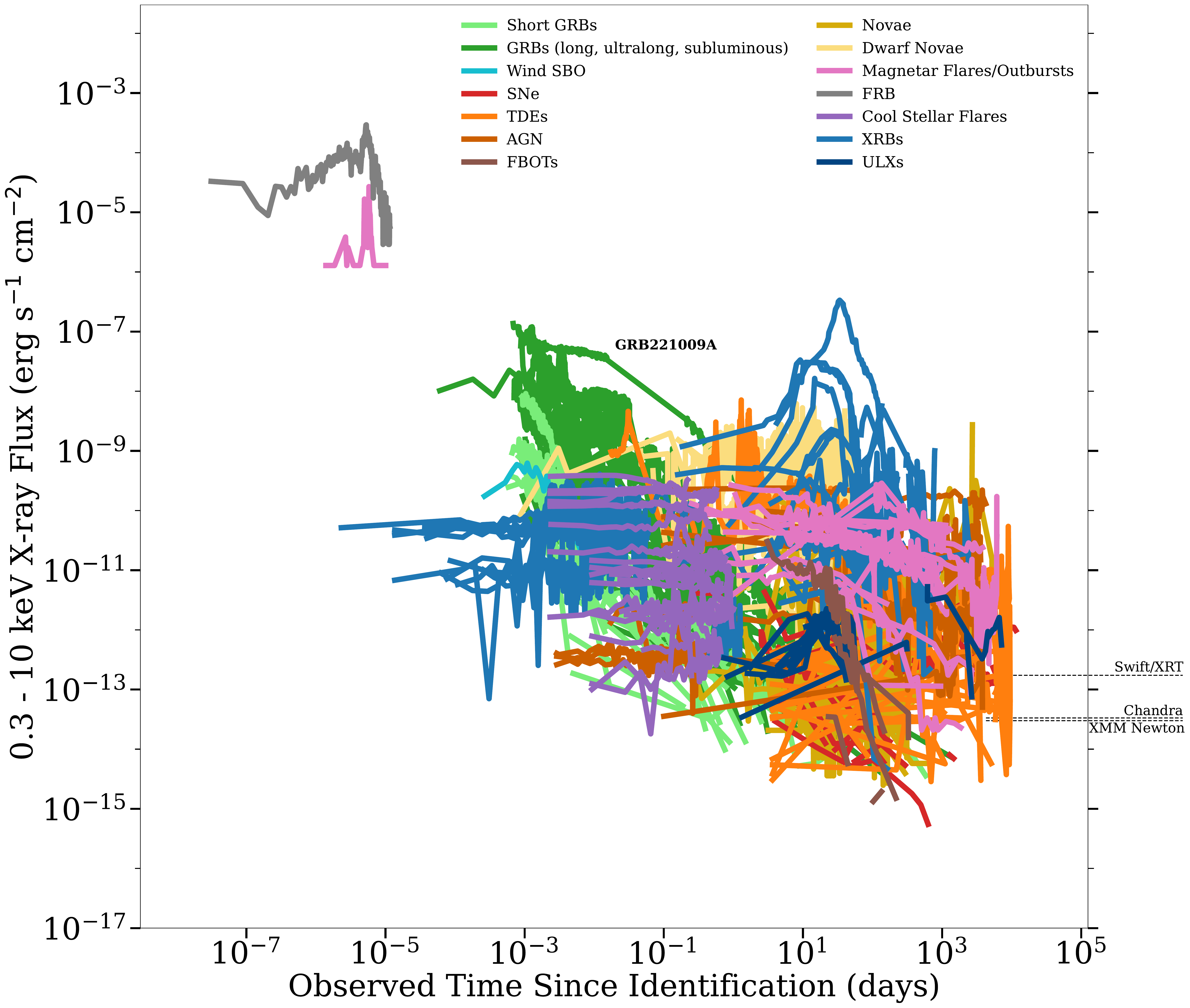}
\caption{The duration-flux phase space of X-ray transient and variable phenomena. To demonstrate the limitations of current observatories, we mark the 0.3-10 keV flux limit for a handful of instruments, assuming a 1ks integration time. \label{fig:DFPS}}
\end{figure*}

While the luminosity X-ray phase space is very instructive for understanding which physical phenomena are poorly captured by current instruments, it does little to reinforce the role that instrument sensitivity plays in determining which phenomena are detected. We show a purely observational duration-flux phase space in Figure \ref{fig:DFPS}.

Future missions will improve substantially on current sensitivity limits. This will open up an innately new area of the low luminosity parameter space, significantly extending the depth out to which known classes of transients can be detected and potentially revealing the existence of yet-unknown intrinsically faint signals. 

\section{The schematic DLPS}
\begin{figure*}[ht!]
\epsscale{1.}
\plotone{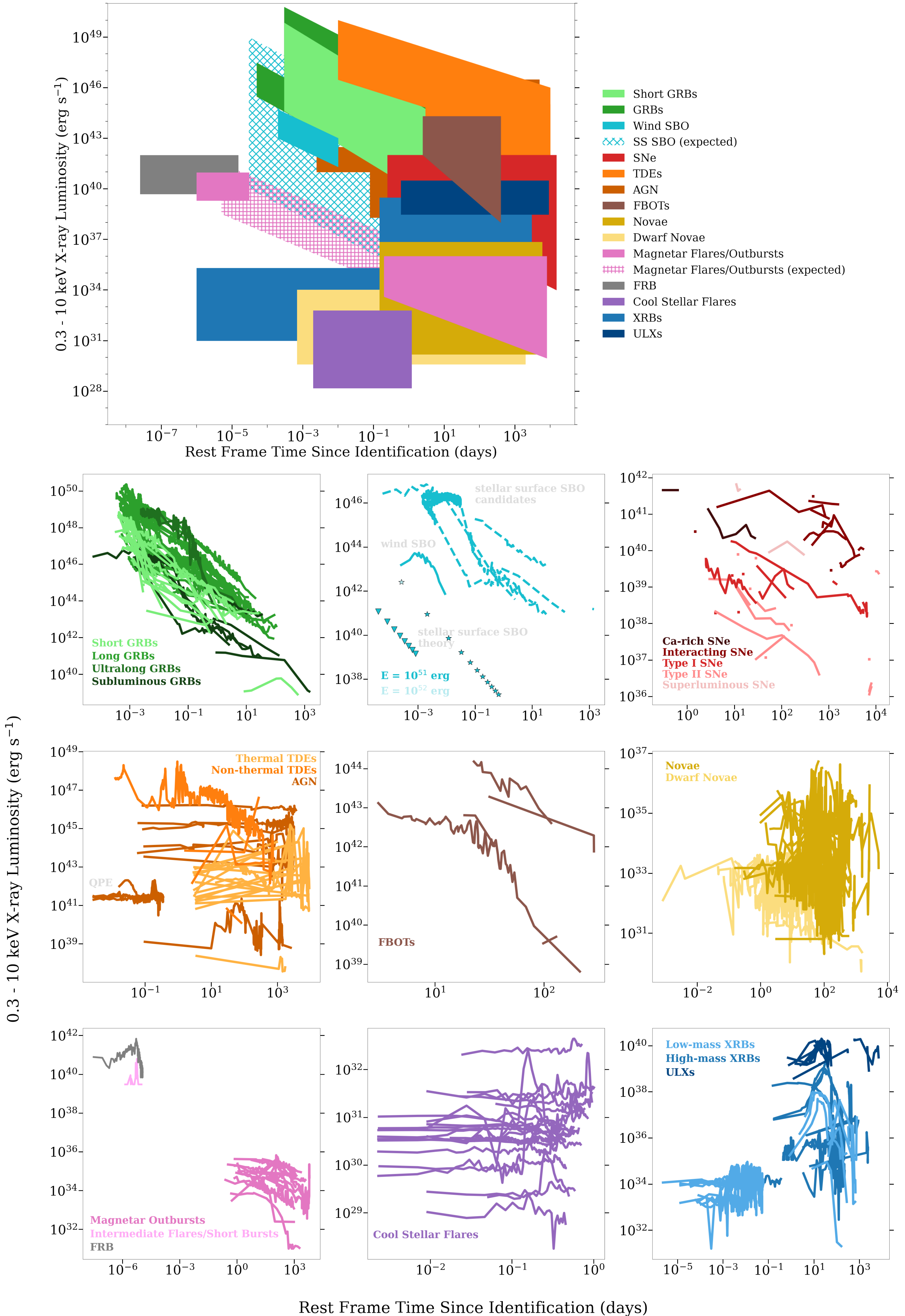}
\caption{\emph{Top:} A schematic illustration of the DLPS. Colored blocks cover the region where light curves of that class exist in the DLPS. The hatched blocks represent where we expect events of a particular class, but have a paucity of observations (for SBOs, we include in this the region of the DLPS where we have \textit{candidate} stellar surface SBOs). \emph{Bottom:} An additional view of the DLPS, which emphasizes the phase space of individual classes of transient rather than their position in the larger transient phase space. We note that the lower panels are all scaled individually in duration and luminosity. \label{fig:schematic}}
\end{figure*}

In addition to the phase space plots that are populated with real light curves, we offer a schematic version of the DLPS in Figure \ref{fig:schematic}. We determine the bounds of the colored blocks based on the specific location of our collected light curves and use these regions as a guide to generate focused subplots that show each (sub)class of transient in greater detail, rather than in the broader context of other signals in the DLPS.

\begin{deluxetable*}{lccccccc}
\tablenum{A1}
\tablecaption{Gamma-ray Bursts\label{tab:GRB}}
\tablewidth{\linewidth}
\tablehead{Name & Type & T$_{90}$ (s) & RA & Dec & z & Distance (kpc) & References}
\startdata
GRB980425A & subluminous & 22.0 & 19:35:03&  -52:50:46 & 0.0085 & $2.7\times10^4$& \citealt{2000ApJ...536..778P, 2004ApJ...608..872K}\\
GRB031203A & subluminous & 30 & 08:02:30 & -39:51:03 & 0.105 & $4.9\times10^5$ & \citealt{2004Natur.430..646S, 2004ApJ...605L.101W}\\
GRB050509B & short & 0.073 & 12:36:18 & +29:01:24& 0.225 & $1.1\times10^6$ & \citealt{2007A-A...469..379E, 2009MNRAS.397.1177E}\\
GRB050724 & short & 3.00 & 16:24:44 & -27:32:28 & 0.258 & $1.3\times10^6$ & \citealt{2007A-A...469..379E, 2009MNRAS.397.1177E}\\
GRB051221A & short & 1.400 & 21:54:49 & +16:53:27 & 0.5465 & $3.2\times10^6$ & \citealt{2007A-A...469..379E, 2009MNRAS.397.1177E}\\
GRB060218A & subluminous & 2100 & 09:09:31 & +33:08:20 & 0.0331 & $1.5 \times10^5$ & \citealt{2007A-A...469..379E, 2009MNRAS.397.1177E}\\
GRB061006 & short & 0.42 & 07:24:08 & -79:11:55 & 0.438 & $2.4\times10^6$ & \citealt{2007A-A...469..379E, 2009MNRAS.397.1177E}\\
GRB061210 & short & 85.0 & 09:38:05 & +15:37:17 & 0.4095 & $2.3\times10^6$ & \citealt{2007A-A...469..379E, 2009MNRAS.397.1177E}\\
GRB061217 & short & 0.210 & 10:41:39&  -21:07:22 & 0.827 & $5.3\times10^6$ & \citealt{2007A-A...469..379E, 2009MNRAS.397.1177E}\\
GRB070714B & short & 3.0 & 03:51:22 & +28:17:51 & 0.923 & $6.1\times10^6$ & \citealt{2007A-A...469..379E, 2009MNRAS.397.1177E}\\
GRB070724A & short & 0.4 & 01:51:14 & -18:35:39 & 0.457 & $2.6\times10^6$ & \citealt{2007A-A...469..379E, 2009MNRAS.397.1177E}\\
GRB071227 & short & 1.8 & 03:52:31 & -55:59:03 & 0.383 & $2.1\times10^6$ & \citealt{2007A-A...469..379E, 2009MNRAS.397.1177E}\\
GRB080905A & short & 1.0 & 19:10:39 & -18:51:55 & 0.1218 & $5.7\times10^5$ & \citealt{2007A-A...469..379E, 2009MNRAS.397.1177E}\\
GRB090510A & short & 0.3 & 22:14:13 & -26:35:51 & 0.903 & $5.9\times10^6$ & \citealt{2007A-A...469..379E, 2009MNRAS.397.1177E}\\
GRB100117A & short & 0.3 & 00:45:05 & -01:35:42& 0.92 & $6.0\times10^6$ & \citealt{2007A-A...469..379E, 2009MNRAS.397.1177E}\\
GRB100316D & subluminous & 292.8 & 07:10:31 & -56:15:20& 0.059 & $2.7\times10^5$ & \citealt{2007A-A...469..379E, 2009MNRAS.397.1177E}\\
GRB100816A & short & 2.9 & 23:26:58 & +26:34:43 & 0.8049 & $5.1\times10^6$ & \citealt{2007A-A...469..379E, 2009MNRAS.397.1177E}\\
GRB101219A & short & 0.6 & 04:58:20 & -02:32:23 & 0.718 & $4.4\times10^6$ & \citealt{2007A-A...469..379E, 2009MNRAS.397.1177E}\\
GRB101225A & ultralong & 1088 & 00:00:47 & +44:36:01& 0.847 & $5.5 \times10^6$ & \citealt{2007A-A...469..379E, 2009MNRAS.397.1177E}\\
GRB141212A & short & 0.30 &02:36:30 & +18:08:49& 0.596 & $3.5\times10^6$ & \citealt{2007A-A...469..379E, 2009MNRAS.397.1177E}\\
GRB141225A & long & 40.24 & 09:15:07 & +33:47:31& 0.915 & $6.0 \times 10^6$ & \citealt{2007A-A...469..379E, 2009MNRAS.397.1177E}\\
GRB150101B & short & 0.08 &12:32:05 & -10:56:01 & 0.134 & $6.4\times10^5$ & \citealt{2007A-A...469..379E, 2009MNRAS.397.1177E}\\
GRB150323A & long & 149.6 & 08:32:43 & +45:27:53& 0.593 & $3.5 \times 10^6$ & \citealt{2007A-A...469..379E, 2009MNRAS.397.1177E}\\
GRB150514A & long & 10.8 &04:59:30 & -60:58:07 & 0.807 & $5.1 \times 10^6$ & \citealt{2007A-A...469..379E, 2009MNRAS.397.1177E}\\
GRB150518A & subluminous & -- & 15:36:48 & +16:19:47 & 0.256 & $1.3\times10^6$ & \citealt{2007A-A...469..379E, 2009MNRAS.397.1177E}\\
GRB150727A & long & 88 & 13:35:53 & -18:19:32& 0.313 & $1.6 \times 10^6$ & \citealt{2007A-A...469..379E, 2009MNRAS.397.1177E}\\
GRB150818A & long & 123.3 & 15:21:25 & +68:20:31& 0.282 & $1.5 \times 10^6$ & \citealt{2007A-A...469..379E, 2009MNRAS.397.1177E}\\
GRB150821A & long & 172.1 & 22:47:39 & -57:53:38 & 0.755 & $4.7 \times 10^6$ & \citealt{2007A-A...469..379E, 2009MNRAS.397.1177E}\\
GRB151027A & long & 129.69 & 18:09:57 & +61:21:12& 0.81 & $5.2\times 10^6$ & \citealt{2007A-A...469..379E, 2009MNRAS.397.1177E}\\
GRB160131A & long & 325 & 05:12:40 & -07:03:00& 0.972 & $6.5 \times 10^6$ & \citealt{2007A-A...469..379E, 2009MNRAS.397.1177E}\\
GRB160314A & long & 8.73 & 07:31:10 & +16:59:57& 0.726 & $4.9\times 10^6$ & \citealt{2007A-A...469..379E, 2009MNRAS.397.1177E}\\
GRB160425A & long & 304.58 & 18:41:19 & -54:21:36& 0.555 & $3.2\times 10^6$ & \citealt{2007A-A...469..379E, 2009MNRAS.397.1177E}\\
GRB160623A & long & 13.5 & 21:01:12 & +42:13:15& 0.367 & $2.0\times 10^6$ & \citealt{2007A-A...469..379E, 2009MNRAS.397.1177E}\\
GRB160624A & short & 0.2 & 22:00:46 & +29:38:38 & 0.483 & $2.7\times10^6$ & \citealt{2007A-A...469..379E, 2009MNRAS.397.1177E}\\
GRB160804A & long & 130 & 14:46:31 & +09:59:56& 0.736 & $4.6 \times 10^3$ & \citealt{2007A-A...469..379E, 2009MNRAS.397.1177E}\\
GRB160821B & short & 0.48 & 18:39:55 & +62:23:30 & 0.16 & $7.7\times10^5$ & \citealt{2007A-A...469..379E, 2009MNRAS.397.1177E}\\
GRB161129A & long & 35.53 & 21:04:55 & +32:08:05& 0.645 & $3.9 \times 10^6$ & \citealt{2007A-A...469..379E, 2009MNRAS.397.1177E}\\
GRB161219B & long & 6.94 & 06:06:5 & -26:47:30& 0.1475 & $7.1 \times 10^5$ & \citealt{2007A-A...469..379E, 2009MNRAS.397.1177E}\\
GRB170519A & long & 216.4 & 10:53:42 & +25:22:27& 0.818 & $5.2\times 10^6$ & \citealt{2007A-A...469..379E, 2009MNRAS.397.1177E}\\
GRB170607A & long & 23.0 & 00:29:28 & +09:14:36& 0.557 & $3.3\times10^6$ & \citealt{2007A-A...469..379E, 2009MNRAS.397.1177E}\\
GRB170714A & ultralong & 1000 & 02:17:24 & +01:59:29 & 0.793 & $5.0\times10^6$ & \citealt{2007A-A...469..379E, 2009MNRAS.397.1177E}\\
GRB170817A & short & 2.0 & 13:09:48 & -23:22:53 & 0.0099 & $4.3\times10^4$ &  \citealt{2019ApJ...886L..17H, 2020RNAAS...4...68H}\\ 
GRB171010A & long & 70.3 &04:26:19 & -10:27:48 & 0.3285 & $1.7\times 10^6$ & \citealt{2007A-A...469..379E, 2009MNRAS.397.1177E}\\
GRB171205A & subluminous & 189.4 & 11:09:39 & -12:35:19 & 0.0368 & $1.6\times10^5$ & \citealt{2007A-A...469..379E, 2009MNRAS.397.1177E}\\
GRB180404A & long & 35.2 &05:34:12 & -37:10:05 & 1.000 & $6.7\times10^6$ & \citealt{2007A-A...469..379E, 2009MNRAS.397.1177E}\\
GRB180703A & long & 20.9 &00:24:28 & -67:18:18 & 0.6678 & $4.0\times10^6$ & \citealt{2007A-A...469..379E, 2009MNRAS.397.1177E}\\
GRB180720B & long & 51.1 & 00:02:07 & -02:55:08 & 0.654 & $4.0 \times10^6$ & \citealt{2007A-A...469..379E, 2009MNRAS.397.1177E}\\
GRB180728A & long & 8.68 & 16:54:16 & -54:02:40& 0.117 & $5.5\times10^5$ & \citealt{2007A-A...469..379E, 2009MNRAS.397.1177E}\\ 
GRB190114C & long & 361.5 & 03:38:01 & -26:56:48 & 0.425 & $2.4 \times 10^6$ & \citealt{2007A-A...469..379E, 2009MNRAS.397.1177E}\\
GRB190829A & long & 63 & 02:58:10 & -08:57:30& 0.0785 & $3.6\times10^5$ & \citealt{2007A-A...469..379E, 2009MNRAS.397.1177E}\\
GRB191019A & long & 64.35 &22:40:06 & -17:19:41 & 0.248 & $1.3\times10^6$ & \citealt{2007A-A...469..379E, 2009MNRAS.397.1177E}\\ \label{GRBtable}
GRB221009A & long & 327 & 19:12:50 & +19:43:48 & 0.151 & $7.2 \times 10^5$ & \citealt{2007A-A...469..379E, 2009MNRAS.397.1177E, GCN32748}
\enddata
\end{deluxetable*}

\begin{deluxetable*}{lccccc}
\tablenum{A2}
\tablecaption{Shock Breakouts\label{tab:SBO}}
\tablewidth{\linewidth}
\tablehead{Name & Type & RA & Dec & Distance (kpc) & References}
\startdata
GRB980425A & stellar surface & 19:35:03 & -52:50.46 & $3.7\times10^4$ & \citealt{2000ApJ...536..778P, 2004ApJ...608..872K}\\
GRB031203A & stellar surface & 08:02:30 & -39:51:03 & $4.9\times10^5$ & \citealt{2004Natur.430..646S, 2004ApJ...605L.101W}\\
GRB060218A & stellar surface & 03:21:40 & +16:52:02 & $1.5 \times10^5$ & \citealt{2007A-A...469..379E, 2009MNRAS.397.1177E}\\
SN2008D (GRB080109A) & wind & 09:09:31 & +33:08:20 & $2.7 \times 10^4$ & \citealt{2008Natur.453..469S}\\
GRB100316D & stellar surface & 07:10:31 & -56:15:20 & $2.7\times10^5$ & \citealt{2007A-A...469..379E, 2009MNRAS.397.1177E}\\
GRB150518A & stellar surface & 15:36:48 & +16:19:47& $1.3\times10^6$ & \citealt{2007A-A...469..379E, 2009MNRAS.397.1177E}\\
GRB171205A & stellar surface & 11:09:39 & -12:35:19& $1.6\times10^5$ & \citealt{2007A-A...469..379E, 2009MNRAS.397.1177E}\\
\enddata
\tablecomments{Subluminous GRBs are considered candidates for stellar surface shock breakouts. We include them here under that assumption.}
\end{deluxetable*}

\begin{deluxetable*}{lccccc}
\tablenum{A3}
\tablecaption{Supernovae\label{tab:SNe}}
\tablewidth{\linewidth}
\tablehead{Name & Type & RA & Dec & Distance (kpc) & References}
\startdata
SN1978K & II &03:17:39  & -66:33:03 & $4.5 \times 10^3$ & Raffaella Margutti, Private Communication\\
SN1981K & II & 12:18:59  & +47:19:31& $7.2 \times 10^3$ & \citealt{2007CBET..828....1I}\\
SN1987A & IIpec & 05:35:28  & -69:16:11&50 &\citealt{2006A-A...460..811H, 2008ApJ...676..361H, 2010A-A...515A...5S}\\
SN1993J & IIb &09:55:25  & +69:01:14 & $2.6 \times 10^3$ & \citealt{2009ApJ...699..388C}\\
SN1995N & IIn & 14:49:28  & -10:10:14& $2.4 \times 10^4$ & \citealt{2005MNRAS.364.1419Z}\\
SN1996cr & IIn &14:13:10  & -65:20:45 & $3.8 \times10^3$ & \citealt{2008ApJ...688.1210B}\\
SN1998bw & Ib/c & 19:35:03  & -52:50:46& $3.8 \times 10^4$ & \citealt{2004ApJ...608..872K}\\
SN1999em & IIP & 04:41:27  & -02:51:45& $7.8  \times  10^3$ & \citealt{2002ApJ...572..932P}\\
SN1999gi & IIP & 10:18:17  & +41:26:28& $8.7 \times 10^3$ & \citealt{2001ApJ...556L..25S}\\
SN2001ig & II & 22:57:31  & -41:02:26 & $1.1 \times 10^4$ & \citealt{2002IAUC.7913....1S}\\
SN2002ap & Ib/c & 01:36:24  & +15:45:13 & $1.0 \times 10^4$ & \citealt{2004A-A...413..107S}\\
SN2003bg & Ic/pec & 04:10:59  & -31:24:50& $1.9 \times 10^4$ & \citealt{2006ApJ...651.1005S}\\
SN2004et & II &20:35:25  & +60:07:18 & $5.5 \times  10^3$ & \citealt{2007MNRAS.381..280M}\\
SN2005ip & IIn &09:32:06  & +08:26:44 & $3.0 \times 10^4$ & \citealt{2007ATel.1004....1I}\\
SN2005kd & IIn & 04:03:17  & +71:43:19& $6.4 \times 10^4$ & \citealt{2007ATel..981....1I, 2007ATel.1023....1P}\\
SN2006bp & IIP & 11:53:56  & +52:21:09& $1.5 \times 10^4$ & \citealt{2007ApJ...664..435I}\\
SN2006jc & Ibc & 09:17:21  & +41:54:33& $2.4 \times 10^4$ & \citealt{2008ApJ...674L..85I}\\
SN2006jd & IIb/IIn & 08:02:07  & +00:48:32& $7.9 \times 10^4$ & \citealt{2007ATel.1290....1I}\\
SN2007pk & IIn & 01:31:47  & +33:36:54& $7.1 \times 10^4$ & \citealt{2007ATel.1284....1I} \\
SN2008M & II & 06:21:41  & -59:43:45& $3.7 \times 10^4$ & \citealt{2010ATel.2478....1I}\\
SN2008ax & IIb & 12:30:41  & +41:38:16& $8.0 \times 10^3$ & \citealt{2009ApJ...704L.118R}\\
SN2008ij & II &18:19:52  & +74:33:55& $2.1 \times 10^4$ & \citealt{2009ATel.1918....1I}\\
SN2009gj & IIb & 00:30:29  & -33:12:56& $1.8 \times 10^4$ & \citealt{2009ATel.2111....1I}\\
SN2009mk & IIb & 00:06:21  & -41:29:00& $2.1 \times 10^4$ & \citealt{2010ATel.2389....1R}\\
SN2010F & II & 10:05:21  & -34:13:21& $3.9 \times 10^4$ & \citealt{2010ATel.2618....1R}\\
SN2010jl & IIn & 09:42:53  & +09:29:42 & $4.9 \times 10^4$ & \citealt{2010ATel.3012....1I, Chandra_2015}\\
SN2011dh & IIb & 13:30:05  & +47:10:11& $7.3 \times 10^3$ & \citealt{2012ApJ...752...78S}\\
SN2011ja & IIP & 13:05:11  & -49:31:27& $3.0 \times 10^3$ & \citealt{2013ApJ...774...30C}\\
SN2013by & IIL/IIn & 16:59:02  & -60:11:42& $1.5 \times 10^4$ & \citealt{2013ATel.5106....1M}\\
SN2013ej & IIP/IIL & 01:36:48  & +15:45:31& $9.6 \times 10^3$ & \citealt{2016ApJ...817...22C}\\
SN2014C & Ib/IIn & 22:37:06  & +34:24:32&  $1.5\times10^4$ & \citealt{2022arXiv220600842B}\\
SN2018gk & IIb/SL & 16:35:54  & +40:01:58 & $1.4 \times 10^5$ & \citealt{2021MNRAS.503.3472B}\\
SN2018bsz\tablenotemark{a} & I/SL & 16:09:39 & -32:03:46 &  $1.1 \times 10^5$ & Matthews et al. in prep\\
SN2019ehk & Ca-rich & 12:22:56  & +15:49:34 & $1.6 \times 10^4$ & \citealt{2020ApJ...898..166J}\\
SN2021gno & Ca-rich & 12:12:10  & +13:14:57& $3.05\times 10^4$ & \citealt{2022ApJ...932...58J}
\enddata
\tablecomments{Type ``SL'' denotes superluminous supernovae. Additional X-ray SNe observations may be, or may become, available in SNaX (\href{https://kronos.uchicago.edu/snax/}{kronos.uchicago.edu/snax}; \citealt{2017AJ....153..246R, 2020RNAAS...4..195N}), which is a moderated database serving as a repository for user-uploaded X-ray observations of supernovae.}
\tablenotetext{a}{These data are not shared in the GitHub repository.}
\end{deluxetable*}

\begin{deluxetable*}{lcccccc}
\tablenum{A4}
\tablecaption{Tidal Disruption Events and Active Galactic Nuclei\label{tab:TDE}}
\tablewidth{\linewidth}
\tablehead{Name & Type & RA & Dec & Distance (kpc) & References}
\startdata
PKS 2155-304 & AGN & 21:58:52. & -30:13:32 & $5.4 \times 10^5$ & \citealt{2018ApJ...852...37A}\\
3C 273 & AGN & 12:29:07 & +02:03:09 & $7.6 \times 10^5$ & \citealt{2018ApJ...852...37A}\\
NGC 4395 & AGN & 12:25:49 & +33:32:49 & $4.7 \times 10^3$ & \citealt{2018ApJ...852...37A}\\
3C 279 & AGN & 12:56:11 & -05:47:22 & $3.1 \times 10^6$& \citealt{2018ApJ...852...37A}\\
3C 345 & AGN & 16:42:59 & +39:48:37 & $3.5 \times 10^6$&\citealt{2018ApJ...852...37A}\\
MKN 335 & AGN & 00:06:20 & +20:12:11 & $1.1 \times 10^5$&\citealt{2018ApJ...852...37A}\\
CGC 229-10 (Zw 299-015) & AGN & 16:41:09 & +61:19:35 & $8.7 \times 10^4$&\citealt{2018ApJ...852...37A}\\
PS10jh & thermal TDE & 16:09:28 & +53:40:23& $8.2 \times 10^5$ & \citealt{2017ApJ...838..149A}\\
ASASSN-14ae & thermal TDE &11:08:40 & +34:05:52 & $2.0 \times 10^5$ & \citealt{2017ApJ...838..149A}\\
ASASSN-14li & thermal TDE & 12:48:15 & +17:46:26& $9.0 \times 10^4$ & \citealt{2015Natur.526..542M, 2017MNRAS.466.4904B}\\
 &  & & & & \citealt{2017ApJ...838..149A, 2018MNRAS.475.4011B}\\
ASASSN-15oi & thermal TDE &20:39:09 & -30:45:21 &$2.2 \times 10^5$ & \citealt{2017ApJ...838..149A, 2018MNRAS.480.5689H}\\
Swift 1644+57 & non-thermal TDE & 16:44:49 & +57:34:51 & $1.9 \times 10^6$ & \citealt{2016ApJ...817..103M, 2017ApJ...838..149A}\\
ASASSN-19bt & non-thermal TDE & 07:00:11 & -66:02:25& $1.15 \times 10^5$ & \citealt{2019ApJ...883..111H}\\
Swift J2058.4+0516\tablenotemark{a} & non-thermal TDE &20:58:20 & +05:13:32 &$1 \times 10^7$ & \citealt{2017ApJ...838..149A}\\
SDSS J131122.15-012345.6 & thermal TDE &13:11:22 & -01:23:46 & $9.0\times10^5$ & \citealt{2017ApJ...838..149A}\\
SDSS J132341.97+482701.3 & thermal TDE & 13:23:42 & +48:27:01 & $4.0\times10^5$ & \citealt{2017ApJ...838..149A} \\
SDSS J1201+3003 & thermal TDE & 12:01:36 & +30:03:06& $7.1 \times 10^5$ & \citealt{2017ApJ...838..149A}\\
WINGS J1348 & thermal TDE & 13:48:51 & +26:35:06 & $2.8 \times 10^5$ & \citealt{2017ApJ...838..149A}\\
RBS 1032 & thermal TDE & 11:47:27 & +49:42:57& $1.1\times10^5$ & \citealt{2017ApJ...838..149A}\\
3XMM J1521+0749 & thermal TDE & 11:47:27& +49:42:58& $8.9\times 10^5$ &  \citealt{2017ApJ...838..149A}\\
GSN 069 & AGN/QPE & 01:19:09 & -34:11:30 & $7.86 \times 10^4$ & \citealt{2019Natur.573..381M}\\
2MASX J0249 & thermal TDE & 02:49:17 & −04:12:52 & $8.0\times10^4$ &\citealt{2017ApJ...838..149A}\\
IGR J17361-4441 & thermal TDE & 17:36:17 & −44:44:06 & $1.8\times10^5$ & \citealt{2017ApJ...838..149A}\\
NGC 247 & thermal TDE & 00:47:09 & −20:45:37 & 2240 & \citealt{2017ApJ...838..149A}\\
OGLE 16aaa & thermal TDE & 01:07:21 & −64:16:21 & $8.1\times10^5 $& \citealt{2017ApJ...838..149A}\\
PTF-10iya & thermal TDE & 14:38:41 & +37:39:33 & $1.2\times10^6$ &\citealt{2017ApJ...838..149A}\\
XMMSL1 J0740-85 & thermal TDE & 07:40:08 & −85:39:31 & $7.4\times10$$^4$ & \citealt{2017ApJ...838..149A}\\
\enddata
\tablenotetext{a}{Though Swift J2058.4+0516 is at $z \sim 1$, we include its light curve anyway due to the relative paucity of non-thermal TDE observations and the uncertainty on its distance estimate.}
\end{deluxetable*}

\begin{deluxetable*}{lcccc}
\tablenum{A5}
\tablecaption{Fast Blue Optical Transients\label{tab:FBOT}}
\tablewidth{\linewidth}
\tablehead{Name & RA & Dec & Distance (kpc) & References}
\startdata
CSS161010 & 04:58:34&  -08:18:04 & $1.5 \times 10^5$ & \citealt{2020ApJ...895L..23C}\\
AT2018cow & 16:16:00 & +22:16:05 & $6.0 \times 10^4$ & \citealt{Margutti_2019}\\
AT2020xnd & 22:20:02 & -02:50:25 & $1.2 \times 10^6$ & \citealt{2022ApJ...926..112B} \\
AT2020mrf & 15:47:54 & +44:29:07 & $6.37\times 10^5$ & \citealt{2022ApJ...934..104Y}\\
AT2022tsd & 03:20:11 & +08:44:56 & $1.3 \times10^6$ & \citealt{2022TNSAN.207....1S, 2022TNSAN.218....1M, 2023TNSAN.159....1M}\\
\enddata
\end{deluxetable*}

\begin{deluxetable*}{lccccc}
\tablenum{A6}
\tablecaption{Cataclysmic Variables \label{tab:novae}\label{tab:CVs}}
\tablewidth{\linewidth}
\tablehead{Name & Type & RA & Dec & Distance (kpc) & References}
\startdata
V838 Her & Nova&18:46:32 +& 12:14:01 & 3.4 & \citealt{2008ApJ...677.1248M}\\
V1974 Cyg & Nova& 20:30:32 & +52:37:51 & 1.9 & \citealt{2008ApJ...677.1248M}\\
V351 Pup & Nova& 08:11:38 & -35:07:30& 4.7 & \citealt{2008ApJ...677.1248M}\\
V382 Vel & Nova& 10:44:48 & -52:25:31& 1.7 & \citealt{2008ApJ...677.1248M}\\
N LMC 2000\tablenotemark{a} & Nova& 05:25:02 & -70:14:17 & 55 & \citealt{2008ApJ...677.1248M}\\
V4633 Sgr & Nova& 18:21:40 & -27:31:37& 8.9 & \citealt{2008ApJ...677.1248M}\\
V5116 Sgr & Nova& 18:17:51 & -30:26:31 & 11.3 & \citealt{2008ApJ...677.1248M}\\
V1663 Aql & Nova& 19:05:12 & +05:14:12 & 5.5 & \citealt{2008ApJ...677.1248M}\\
V477 Sct & Nova& 18:38:43 & -12:16:16 & 11 & \citealt{2008ApJ...677.1248M}\\
V382 Nor & Nova& 16:19:45 & -51:34:53 & 13.8 & \citealt{2008ApJ...677.1248M}\\
RS Oph & Nova& 17:50:13 & -06:42:28 & 1.6 & \citealt{2020AdSpR..66.1169P}\\
V2362 Cyg & Nova& 21:11:32 & +44:48:04 & 7.2 - 15.8 & \citealt{2009NewA...14....4P, 2020AdSpR..66.1169P}\\
V1280 Sco & Nova& 16:57:41 & -32:20:36& 1.6 & \citealt{2008A_A...487..223C, 2020AdSpR..66.1169P}\\
V1281 Sco & Nova& 16:56:59 & -35:21:50& 25.9 & \citealt{2017arXiv170304087K, 2020AdSpR..66.1169P}\\
V458 Vul & Nova& 19:54:25 & +20:52:53 & 8.5 & \citealt{2020AdSpR..66.1169P}\\
V597 Pup & Nova& 08:16:18 & -34:15:25& 3 & \citealt{2020A_A...639A..17W, 2020AdSpR..66.1169P}\\
V2468 Cyg & Nova& 19:58:34 & +29:52:12& 5.6& \citealt{2015AJ....149..136R, 2020AdSpR..66.1169P}\\
V2491 Cyg & Nova& 19:43:02 & +32:19:14& 10.5 - 14& \citealt{2011AA...530A..70D, 2020AdSpR..66.1169P}\\
HV Cet (CSS081007) & Nova& 03:05:59 & +05:47:16& 4.45 & \citealt{2020AdSpR..66.1169P}\\
LMC 2009a & Nova& 05:04:44 & -66:40:12 & 50 & \citealt{2020AdSpR..66.1169P}\\
V2672 Oph & Nova& 17:38:20 & -26:44:14& 19 & \citealt{2011MNRAS.410..525M, 2020AdSpR..66.1169P}\\
KT Eri & Nova& 04:47:54 & -10:10:43& 6.3 & \citealt{2013MNRAS.433.2657R, 2020AdSpR..66.1169P}\\
U Sco & Nova& 16:22:31 & -17:52:43& 12 & \citealt{2010ApJS..187..275S, 2020AdSpR..66.1169P}\\
V407 Cyg & Nova& 21:02:10 & +45:46:33 & 2.7 & \citealt{2020AdSpR..66.1169P}\\
T Pyx & Nova& 09:04:42 & -32:22:48& 3.185 & \citealt{2018MNRAS.481.3033S, 2020AdSpR..66.1169P}\\
LMC 2012 & Nova& 04:54:57 & -70:26:56& 50 & \citealt{2020AdSpR..66.1169P}\\
V959 Mon & Nova& 06:39:39 & +05:53:53 & 1.4 & \citealt{2020AdSpR..66.1169P, 2020ApJ...905..114L}\\
SMC 2012 & Nova& 00:32:34 & -74:20:15 & 61 & \citealt{2020AdSpR..66.1169P}\\
V339 Del & Nova& 20:23:31 & +20:46:04 & 2.1 & \citealt{2020AdSpR..66.1169P, 2020ApJ...905..114L}\\
V1369 Cen & Nova& 13:54:45 & -59:09:04 & 2.0 & \citealt{2020AdSpR..66.1169P, 2020ApJ...905..114L}\\
V745 Sco & Nova& 17:55:22 & -33:14:59 & 7.8 & \citealt{2010ApJS..187..275S, 2020AdSpR..66.1169P}\\
V1534 Sco & Nova& 17:15:47 & -31:28:30 & 8.8 & \citealt{2018ApJS..237....4H, 2020AdSpR..66.1169P}\\
V1535 Sco & Nova& 17:03:26 & -35:04:18 & 8.5 & \citealt{2017ApJ...842...73L, 2020AdSpR..66.1169P}\\
V5668 Sgr & Nova& 18:37:40 & -29:04:03 & 2.0 & \citealt{2020AdSpR..66.1169P, 2020ApJ...905..114L}\\
LMC 1968-12a & Nova& 05:09:58 & -71:39:53 & 50 & \citealt{2020AdSpR..66.1169P}\\
V407 Lup & Nova& 15:29:02 & -44:49:41 & $\sim10$ & \citealt{2018MNRAS.480..572A, 2020AdSpR..66.1169P}\\
SMCN 2016-10a & Nova& 01:06:03 & -74:47:16 & 61 & \citealt{2020AdSpR..66.1169P}\\
V549 Vel & Nova & 08:50:30 & -47:45:28 & 0.560 & \citealt{2020AdSpR..66.1169P, 2020ApJ...905..114L}\\
SS Cyg & Dwarf Nova & 21:42:43 & +43:35:10 & 0.115 & \citealt{2003MNRAS.345...49W, 2004ApJ...601.1100M, 2020MNRAS.494.3799P}\\
GW Lib & Dwarf Nova & 15:19:55 & -25:00:25 & 0.113 & \citealt{2009MNRAS.399.1576B, 2018A-A...611A..13N, 2020MNRAS.494.3799P}\\
SSS J122221.7−311525 & Dwarf Nova & 12:22:22 & −31:15:24 & 0.275 & \citealt{2018A-A...611A..13N}
\enddata
\tablecomments{We include only the dwarf novae with well-observed X-ray brightening during their optical outbursts.}
\tablenotetext{a}{We quote the 55 kpc distance assumed by \citet{2008ApJ...677.1248M} since these light curves are from that paper and presented as luminosity vs. time. Other novae in the LMC are listed with a more recently revised distance \citep[][]{2013Natur.495...76P} as those data were initially presented as flux vs. time.}
\end{deluxetable*}

\begin{deluxetable*}{lccccc}
\tablenum{A7}
\tablecaption{Magnetar Flares/Outbursts + FRBs \label{tab:magnetars}}
\tablewidth{\linewidth}
\tablehead{Name & Type & RA & Dec & Distance (kpc) & References}
\startdata
1E161348-5055 & Outburst & 16:17:33 & -51:02:00 & 3.3 & \citealt{2016ApJ...828L..13R, 2019A-A...626A..19E}\\
SGR 1627-41 & Outburst & 16:35:52 & -47:35:12 & 11 & \citealt{2018MNRAS.474..961C}\\
1E2259+586 & Outburst & 23:01:08 & +58:52:44& 3.2 & \citealt{2018MNRAS.474..961C}\\
XTE J1810-197 & Outburst & 18:09:51 & -19:43:52& 3.5 & \citealt{2018MNRAS.474..961C}\\
SGR 1806-20 & Outburst &18:08:39 & -20:24:40 & 8.7 & \citealt{2018MNRAS.474..961C}\\
CXOU J1647-4552 & Outburst & 16:47:10 & -45:52:17 & 4 & \citealt{2018MNRAS.474..961C}\\
SGR 0501+4516 & Outburst &05:01:08 & +45:16:31 & 1.5 & \citealt{2018MNRAS.474..961C}\\
1E1547.0-5408 & Outburst & 15:50:54 & -54:18:24& 4.5 & \citealt{2018MNRAS.474..961C}\\
SGR 0418+5729 & Outburst & 04:18:34 & +57:32:23& 2 & \citealt{2018MNRAS.474..961C}\\
SGR 1833-0832 & Outburst & 18:33:46 & -08:32:13& 10 & \citealt{2018MNRAS.474..961C}\\
Swift J1822.3-1606 & Outburst & 18:22:18 & -16:04:27& 1.6 & \citealt{2018MNRAS.474..961C}\\
Swift J1834.9-0846 & Outburst &18:34:53 & -08:45:41 & 4.2 & \citealt{2018MNRAS.474..961C}\\
1E1048.1-5937 & Outburst & 10:50:09 & -59:53:20& 9 & \citealt{2018MNRAS.474..961C}\\
SGR J1745-2900 & Outburst &17:45:40 & -29:00:30 & 8.3 & \citealt{2018MNRAS.474..961C}\\
SGR 1935+2154 (FRB 200428)\tablenotemark{a} & FRB & 19:34:56 & +21:53:48& 4.4 & \citealt{2021NatAs...5..378L}\\
SGR 1935+2154 & IF/SB & & & & \citealt{2009PASJ...61..999M, 2020GCN.27661....1S}
\enddata
\tablecomments{As with the other variable classes, one listed object may correspond to multiple light curves within our X-ray phase space. To remain consistent with our discussion in Section \ref{subsec:magnetar}, we categorize magnetar transience as either intermediate flare/short burst (IF/SB in the table) or outburst. Quiescent behavior is shown for the listed outbursts, with L$_x$ taken from \citet{Olausen_2014}.}
\tablenotetext{a}{SGR 1935+2154 is believed to be a fast radio burst X-ray counterpart with a magnetar progenitor. For that reason, we include it with our sample of magnetar flares and outbursts. These data (both the IF/SB and FRB counterpart) are from the same burst forest in April 2020 for direct comparison. We adopt a distance of 4.4 kpc from \citet{2020ApJ...898L..29M}.}
\end{deluxetable*}

\begin{deluxetable*}{lcccc}
\tablenum{A8}
\tablecaption{Cool Stellar Flares\label{tab:mdwarfs}}
\tablewidth{\linewidth}
\tablehead{Name & RA & Dec & Distance (kpc) & References}
\startdata
UY Scl & 00:14:46 & -39:14:36& 0.1372 & \citealt{2015A-A...581A..28P}\\
HD 1165 & 00:16:53 & +81:39:49& 0.0332 & \citealt{2015A-A...581A..28P}\\
HD 14716 & 02:16:04 & -73:50:43& 0.062 & \citealt{2015A-A...581A..28P}\\
CC Eri & 02:34:23 & -43:47:47& 0.0116 & \citealt{2015A-A...581A..28P}\\
CD-53 544 & 02:41:47 & -52:59:52& 0.028 & \citealt{2015A-A...581A..28P}\\
SDSS J033815.04+002926.0 & 03:38:15 & +00:29:26 & 0.7099 & \citealt{2015A-A...581A..28P}\\
V471 Tau & 03:50:25 & +17:14:47 & 0.0441 & \citealt{2015A-A...581A..28P}\\
2MASS J04072181-1210033 & 04:07:22 & -12:10:03& 0.3957 & \citealt{2015A-A...581A..28P}\\
V410 Tau & 04:18:31 & +28:27:16& 0.0982 & \citealt{2015A-A...581A..28P}\\
T Tau & 04:21:59 & +19:32:06& 0.1825 & \citealt{2015A-A...581A..28P}\\
HD 285845 & 04:31:25 & +18:16:17& 0.090 & \citealt{2015A-A...581A..28P}\\
HD 283810 & 04:40:09 & +25:35:33& 0.060 & \citealt{2015A-A...581A..28P}\\
HD 268974 & 05:05:27 & -67:43:14& 0.9174 & \citealt{2015A-A...581A..28P}\\
AB Dor & 05:28:45 & -65:26:55& 0.0152 & \citealt{2015A-A...581A..28P}\\
SV Cam & 06:41:19 & +82:16:02 & 0.088 & \citealt{2015A-A...581A..28P}\\
pi.01 UMa & 08:39:12 & +65:01:15 & 0.0144 & \citealt{2015A-A...581A..28P}\\
2MASS J13141103-1620235 & 13:14:11 & -16:20:24 & 0.5161 & \citealt{2015A-A...581A..28P}\\
1RXS J231628.7+790531 & 23:16:31 & +79:05:36& 0.055 & \citealt{2015A-A...581A..28P}
\enddata
\tablecomments{As with the progenitors of other classes of recurrent outburst, individual flares are shown separately in our X-ray phase space, so some of the objects listed may correspond to a number of unique light curves.}
\end{deluxetable*}

\begin{deluxetable*}{lccccc}
\tablenum{A9}
\tablecaption{X-ray Binary Outbursts and Ultraluminous X-ray Sources\label{tab:XRB-ULX}}
\tablewidth{\linewidth}
\tablehead{Name & Type & RA & Dec &Distance (kpc) & Reference}
\startdata
4U 0352-309 (X Persei) & HMXRB & 03:55:23 & +31:02:45& 1&  \citealt{2007A-A...474..137L}\\
XMMU J004243.6+412519 & ULX &00:42:44 & +41:25:19 & 778 & \citealt{2013Natur.493..187M}\\
RX J0209.6-7427 & HMXRB & 02:09:34 & -74:27:12 & 55 & \citealt{2020MNRAS.494.5350V}\\
PSR J1023+0038\tablenotemark{a} & LMXRB & 10:23:48 & +00:38:41 & 1.37 & \citealt{2015ApJ...806..148B}\\
IGR J01217-7257 (SXP 2.16) & HMXRB & 01:21:41 & −72:57:22 & 62 & \citealt{2017MNRAS.466.1149B,2017MNRAS.470.1971V}\\
SXP 15.6 & HMXRB & 00:48:55 & -73:49:46& 62 & \citealt{2017MNRAS.470.4354V}\\
CG X-1 & ULX & 14:13:12 & -65:20:14& 4200 & \citealt{2019ApJ...877...57Q}\\
M51 ULX-7 & ULX & 13:30:01 & +47:13:44& 8580 & \citealt{2020MNRAS.491.4949V}\\
NGC 925 ULX-3 & ULX & 02:27:20 & +33:34:13 & 9560 & \citealt{2020ApJ...891..153E}\\
Aql X-1\tablenotemark{b} & LMXRB & 19:11:16 & +00:35:06 & $\sim5$ & \citealt{2020MNRAS.493..940L}\\
GX 339-4\tablenotemark{b} & LMXRB & 17:02:49 & -48:47:23& 8 & \citealt{2000MNRAS.312L..49K, 2013MNRAS.428.2500C}\\
MAXI J1659-152 & LMXRB & 16:59:02 & -15:15:29 & 6 & \citealt{2012MNRAS.423.3308J}\\
4U J1907+09 & HMXRB & 19:09:41 & +09:48:25 & 5 & \citealt{2022AA...664A..99F}\\
IGR J16393-4643 & HMXRB & 16:39:06 & -46:42:14 & 12 & \citealt{2022AA...664A..99F}\\
IGR J17503-2636 & HMXRB & 17:50:18 & -26:36:17 & 10 & \citealt{2022AA...664A..99F}\\
IGR J19140+0951 & HMXRB &19:14:04 & +09:52:58 & 2.8 & \citealt{2022AA...664A..99F}\\
Swift J0243.6+6124 & HMXRB & 02:43:40 & +61:26:04 & 7 & \citealt[][]{2018ApJ...863....9W, 2022MNRAS.509.2532C}\\
RX J0520.5-6932 & HMXRB &05:20:31& -69:31:55 & 50 & \citealt[][]{2014AA...567A.129V}\\
SMC X-2 & HMXRB & 00:54:33 & -73:41:01& 62 & \citealt[][]{2017ApJ...834..209L}\\
SMC X-3 & HMXRB &00:52:06& -72:26:04 & 62 & \citealt[][]{2018AA...614A..23K}\\
XMMU J053108.3-690923 & HMXRB & 05:31:08 & -69:09:24 & 50 & \citealt[][]{2018MNRAS.475..220V, 2021AA...647A...8M}\\
\hline
XTE J1859+226 & LMXRB & 18:58:42 & +22:39:29 & 6.3 & \citealt{2003A_A...399..631H, 2008ApJ...683L..51G}\\
GS 2023+338 & LMXRB & 20:24:04 & +33:52:02& 3.5 & \citealt{2000MNRAS.312L..49K}\\
4U 1630-47 & LMXRB & 16:34:02 & -47:23:35 & 10 & \citealt{2000MNRAS.312L..49K}\\
CXOGLB J173617.6−444416 & LMXRB & 17:36:18 & −44:44:17&9.9& \citealt{2012ApJ...756..147M}\\
CXOGLB J173616.9−444409 & LMXRB & 17:36:17 & −44:44:10&9.9& \citealt{2012ApJ...756..147M}\\
CXOGLB J173617.3−444408 & LMXRB & 17:36:17 & −44:44:08 &9.9& \citealt{2012ApJ...756..147M}\\
CXOGLB J173618.1−444359 & LMXRB & 17:36:18 & −44:43:59 &9.9& \citealt{2012ApJ...756..147M}\\
CXOGLB J173617.5−444357 & LMXRB & 17:36:18 & −44:43:57 &9.9& \citealt{2012ApJ...756..147M}\\
IGR J17544-2619 & HMXRB & 17:54:25 & -26:19:53& 3.5, 3.6 & \citealt{2005A-A...441L...1I, 2008ApJ...687.1230S}\\
IGR J08408–4503 & HMXRB & 08:40:48 & -45:03:32 & 2.7 & \citealt{2007A-A...465L..35L}\\
IGR J16479-4514 & HMXRB & 16:48:07 & -45:12:07 & 4.9 & \citealt{2008ApJ...687.1230S}\\
XTE J1739-302 & HMXRB & 17:39:12 & -30:20:38 & 2.7 & \citealt{2008ApJ...687.1230S}\\
IGR J18410-0535 & HMXRB & 18:41:00 & -05:35:46 & 5 & \citealt{2008ApJ...687.1230S}\\
CI Cam \tablenotemark{c} & ULX & 04:19:42 & +55:59:58 & 1-10 & \citealt{2019A-A...622A..93B}
\enddata
\tablecomments{Objects with light curves shown in the X-ray phase space are above the horizontal line. Below the line, we list objects for which we show the quiescent behavior. As with other variable phenomena that show recurrent outbursts and flares, some objects may correspond to a number of unique light curves.}
\tablenotetext{a}{These data are not shared in the GitHub repository.}
\tablenotetext{b}{The light curves \textit{and} quiescent behavior of Aql X-1 and GX 339-4 are shown in the X-ray phase space.}
\tablenotetext{c}{We note that CI Cam is a ULX as long as it is at a distance $>8$ kpc.}
\end{deluxetable*}

\begin{deluxetable*}{lccccc}
\tablenum{A10}
\tablecaption{Unclassified X-ray Sources \label{tab:oddballs}}
\tablewidth{\linewidth}
\tablehead{Name &  RA & Dec & z & Distance (kpc) & Reference}
\startdata
XRT 000519 & 12:25:32 & +13:03:59& 0.23 - 1.5 & $1.62 \times 10^4$ &  \citealt{Jonker_2013}\\ 
 & & & & $1.1 \times 10^6$ & \\
 & & & & $1.11 \times 10^7$ & \\
XRT 110103\tablenotemark{a} & 14:08:29 & -27:03:29 & - & $9.49 \times 10^4$ & \citealt{Glennie_2015}\\
XRT 120830\tablenotemark{a} & 23:52:12 & -46:43:43& - & 0.08 & \citealt{Glennie_2015}\\
Source 1 & 12:42:51 &  +02:38:35 & - & $1.43 \times 10^4$& \citealt{2016Natur.538..356I}\\
Source 2 & 13:25:53 &  -43:05:46 & - &  $3.8 \times 10^3$ & \citealt{2016Natur.538..356I}\\
CDF-S XT1\tablenotemark{b} & 03:32:39 & −27:51:34 & 0.3 - 2.23 & $1.6 \times 10^6$ &\citealt{Bauer_2017}\\
 & & & & $1.81 \times 10^7$ & \\
CDF-S XT2 & 03:32:18 &  -27:52:24 & 0.738 &  $4.68 \times 10^6$ & \citealt{2019Natur.568..198X}\\
EXMM 023135.0-603743 & 02:31:35 & -60:37:43& 0.092 & $4.35\times 10^5$ & \citealt{2020ApJ...898...37N}\\
\enddata
\tablenotetext{a}{These data are digitized \citep[][]{WebPlotDigitizer} and so are not included in the GitHub repository of light curves from this paper.}
\tablenotetext{b}{See also \citet{2022AA...663A.168Q, 2023arXiv230413795Q} for data and analysis.}
\end{deluxetable*}

\clearpage
\bibliography{references.bib}
\bibliographystyle{aasjournal}

\end{document}